\documentclass[preprint2]{aastex}

\newcommand{\elecd}{$n_{\rm e}$}
\newcommand{\te}{$T_{\rm e}$}
\newcommand{\hb}{H$\beta$}
\newcommand{\ha}{H$\alpha$}
\newcommand{\hg}{H$\gamma$}
\newcommand{\hd}{H$\delta$}
\newcommand{\foi}{[O~{\sc i}]}
\newcommand{\foii}{[O~{\sc ii}]}
\newcommand{\foiii}{[O~{\sc iii}]}

\newcommand{\fsii}{[S~{\sc ii}]}
\newcommand{\fsiii}{[S~{\sc iii}]}

\newcommand{\fnii}{[N~{\sc ii}]}

\newcommand{\fariii}{[Ar~{\sc iii}]}
\newcommand{\fariv}{[Ar~{\sc iv}]}
\newcommand{\farv}{[Ar~{\sc v}]}

\newcommand{\fcliii}{[Cl~{\sc iii}]}
\newcommand{\fcliv}{[Cl~{\sc iv}]}
\newcommand{\fneiii}{[Ne~{\sc iii}]}

\newcommand{\nitroi}{N~{\sc i}}
\newcommand{\nii}{N~{\sc ii}}
\newcommand{\niii}{N~{\sc iii}}
\newcommand{\oi}{O~{\sc i}}
\newcommand{\oii}{O~{\sc ii}}

\newcommand{\cii}{C~{\sc ii}}
\newcommand{\ciii}{C~{\sc iii}}

\newcommand{\nei}{Ne~{\sc i}}
\newcommand{\neii}{Ne~{\sc ii}}

\newcommand{\hi}{H\,{\sc i}}
\newcommand{\hii}{H~{\sc ii}}
\newcommand{\hei}{He~{\sc i}}
\newcommand{\heii}{He~{\sc ii}}
\newcommand{\mc}{\multicolumn}
\newcommand{\ph}{\phantom}
\newcommand{\tnm}{\tablenotemark}
\newcommand{\cb}{$c$(H$\beta$)}
\newcommand{\adf}{{\it adf}}
\newcommand{\adfs}{{\it adf}s}
\newcommand{\kms}{km~s$^{-1}$}
\newcommand{\nd}{\nodata}
\newcommand{\msun}{M$_\odot$}
\newcommand{\rsun}{R$_\odot$}

\shorttitle{Binarity and the abundance discrepancy problem}
\shortauthors{Corradi et al.}

\begin{document}


\title{Binarity and the abundance discrepancy problem
\\ in planetary  nebulae}


\author{Romano L.M. Corradi,
Jorge Garc\'\i a-Rojas,
David Jones,
and Pablo Rodr{\'{\i}}guez--Gil
}
\affil{Instituto de Astrof{\'{\i}}sica de Canarias, E-38200 La Laguna, Tenerife, Spain\\
 Departamento de Astrof{\'{\i}}sica, Universidad de La
  Laguna, E-38206 La Laguna, Tenerife, Spain}


\begin{abstract}
The discrepancy between abundances computed using optical
recombination lines (ORLs) and collisionally excited lines (CELs) is a
major unresolved problem in nebular astrophysics.  We show here that
{\it the largest abundance discrepancies are reached in planetary
nebulae with close binary central stars}.  This is illustrated by deep
spectroscopy of three nebulae with a post common-envelope (CE) binary
star. Abell 46 and Ou5 have O$^{2+}$/H$^+$ abundance discrepancy
factors larger than 50, and as high as 300 in the inner regions of
Abell 46. Abell 63 has a smaller discrepancy factor
around 10, but still above the typical values in ionized nebulae.  Our
spectroscopic analysis supports previous conclusions that, in addition
to ``standard'' hot (\te$\sim$10$^4$~K) gas, a colder
(\te$\sim$10$^3$~K) ionized component that is highly enriched in heavy
elements also exists.  These nebulae have low ionized masses, between
10$^{-3}$ and 10$^{-1}$~\msun\ depending on the adopted electron
densities and temperatures.  Since the much more massive red-giant
envelope is expected to be entirely ejected in the CE phase, the
currently observed nebulae would be produced much later, in post--CE
mass loss episodes when the envelope has already dispersed.  These
observations add constraints to the abundance discrepancy
problem. Possible explanations are revised. Some are naturally linked
to binarity, such as for instance high-metallicity nova ejecta, but it
is difficult at this stage to depict an evolutionary scenario
consistent with all the observed properties.  The hypothesis that
these nebulae are the result of tidal destruction, accretion and
ejection of Jupiter-like planets is also introduced.
\end{abstract}


\keywords{planetary nebulae: individual (A~46, A~63, Ou5) -- ISM:
  abundances - binaries: close -- novae, cataclysmic variables -- 
planet--star interactions}

\section{Introduction}

This work deals with two apparently unrelated, main topics   
in the study of planetary nebulae (PNe).

The first one is the role of binary evolution, which is the favoured
explanation of the diverse morphologies displayed by PNe
(e.g. \citealt{s97,bf02}), but which may even be the cause of
their mere existence \citep{mdm06,s06}.

The second issue is the so-called {\it abundance discrepancy
  problem}. It is well known (see e.g. Osterbrock \& Ferland 2006)
that in photoionized nebulae -- both PNe and H~II regions -- optical
recombination lines (ORLs) provide abundance values that are
systematically larger than those obtained using collisionally excited
lines (CELs).  This a long-standing problem in nebular astrophysics,
and has obvious implications on the measurement of the chemical
content of the Universe, often done using CELs from emission regions
and the ISM.  The {\it abundance discrepancy factor} (\adf\,) between
ORLs and CELs is usually between 1.5 and 3 (see
e.g. \citealt{g07,l12,e14}), but in PNe it has a significant tail
extending to much larger values. The nebula with the largest known
\adf\ \citep[$\sim$70,][]{l06} is Hf~2--2, which has a close binary
central star \citep{lutz98}.

In our recent study of the new Galactic PN IPHASXJ211420.0+434136
(Ou5), which also has a binary central star \citep{c14}, we
noticed the unusual strength of recombination lines such as
\cii\,4267~\AA\ and the 4650~\AA\ \oii+\niii\ blend.  This points to
large \adf\ values also in this object, as well as in other
  binary PNe such as Abell~46 \citep{bond80} and Abell~63
  \citep{bond78} in which strong \cii\ emission was detected
  \citep{pb97}.  The short orbital periods, between 8 and 11 hours,
  indicate that the central stars of these PNe have gone through a
  common--envelope (CE) phase in their previous evolution.
Therefore, we decided to obtain deep spectroscopy of these three
nebulae to investigate whether the abundance discrepancy problem in
PNe is in some way related to the binary nature of their central
stars. That this is indeed true is demonstrated by the results
presented in the following.

\section{Observations}

\begin{figure}[!ht]
\epsscale{0.7755}
\plotone{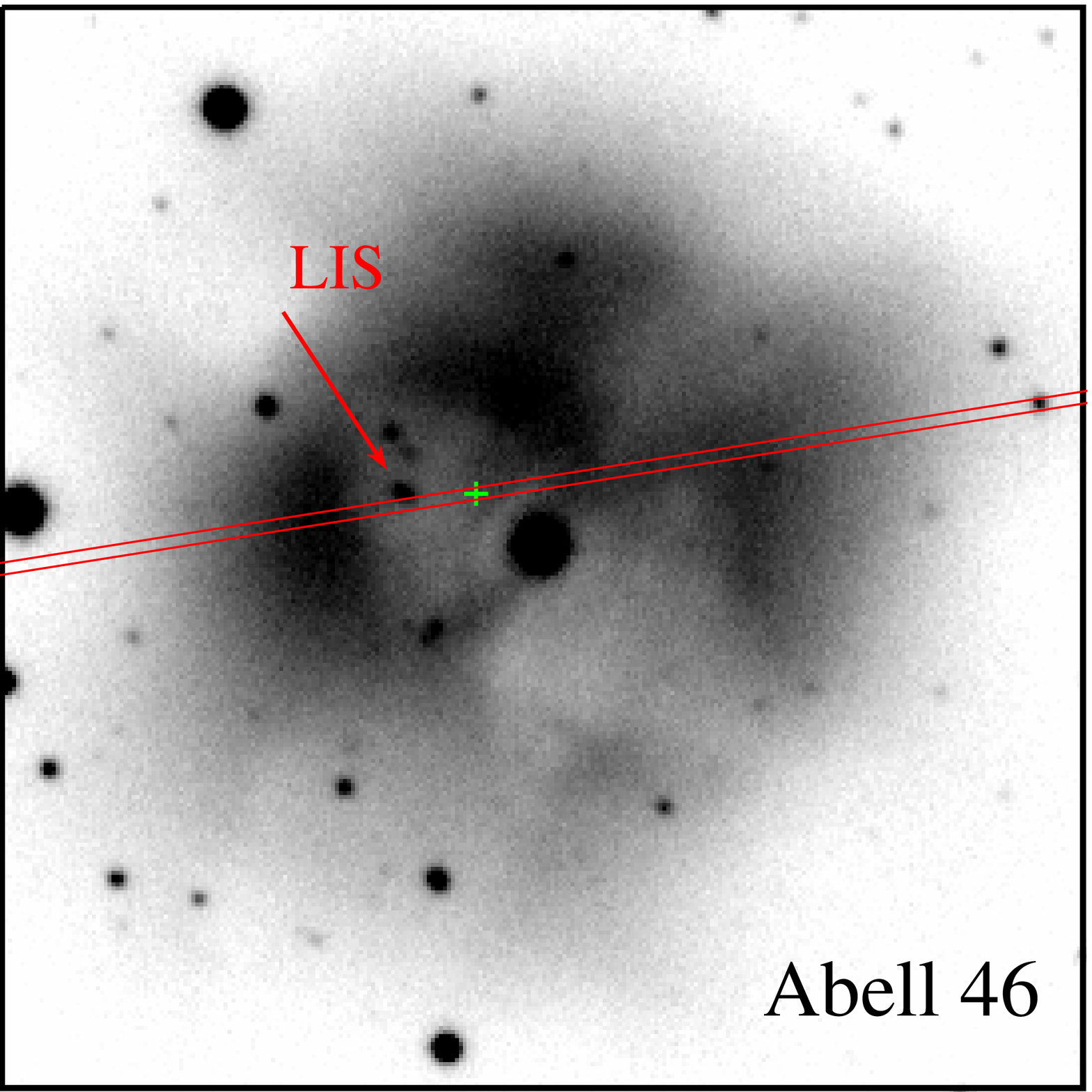}
\plotone{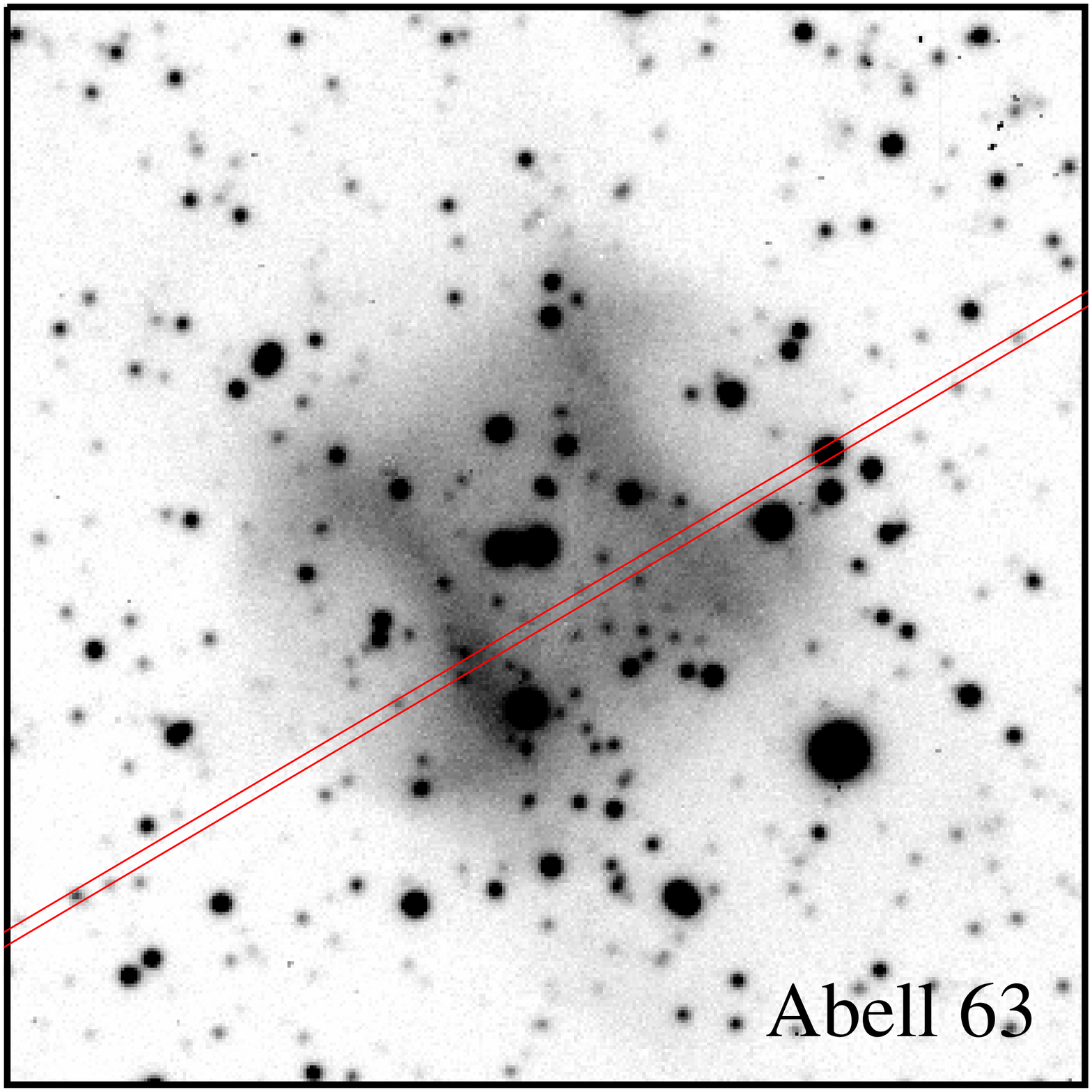}
\plotone{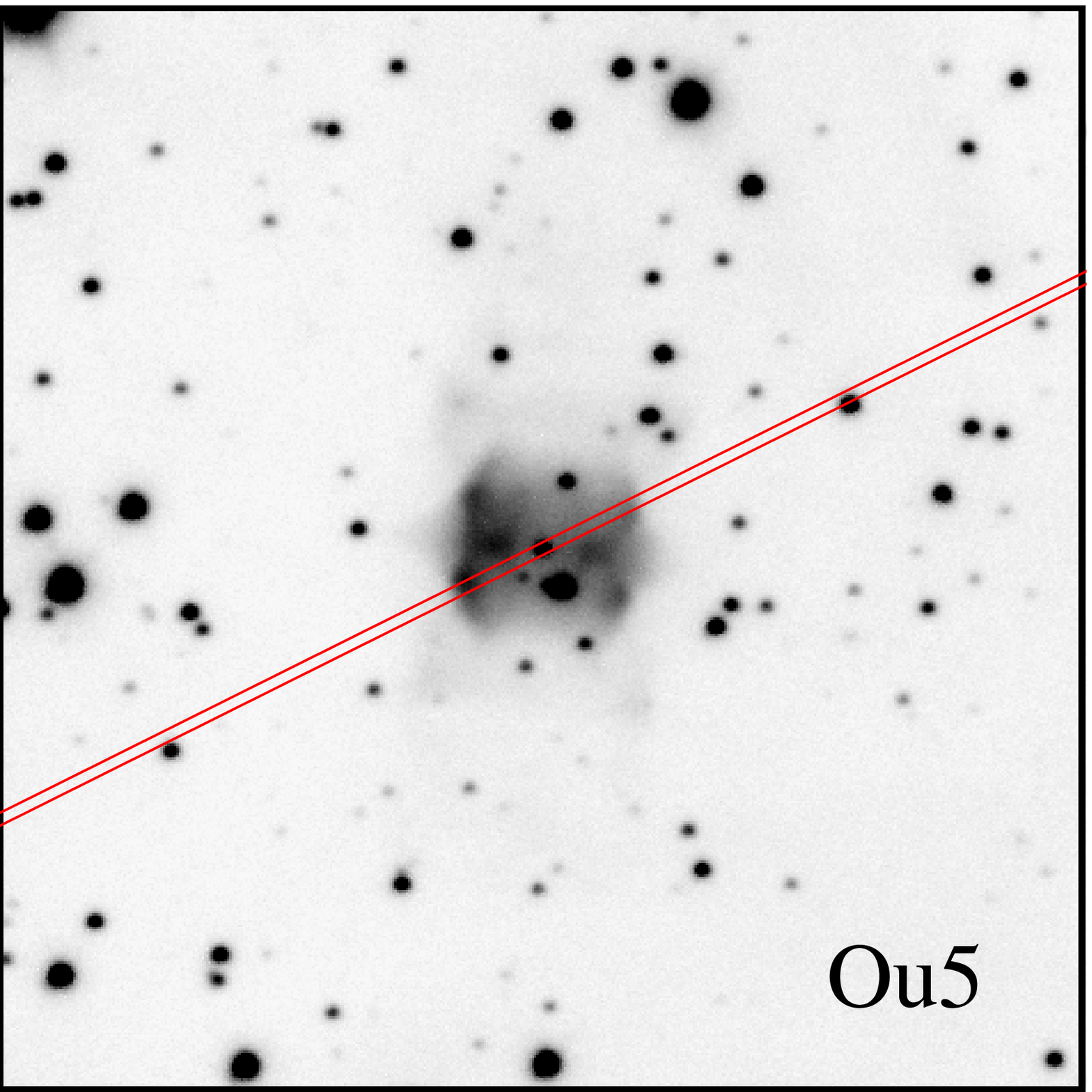}
\caption{\ha+\nii\ images of the nebulae. The field of view is
  90\arcsec$\times$90\arcsec\ in all frames.  The location of the 
  long slit is shown. In A~46, the green cross indicates where 
  emission of recombination lines peaks along the
  slit. The position of the LIS identified in the nebula is also indicated.}
\label{F-neb}
\end{figure}

\ha+\nii\ images of A~46 and A~63 were retrieved from the archives of
the Isaac Newton Group of Telescopes and ESO.  The image of A~46 is an
1-hr exposure from the 2.5m~Isaac Newton Telescope on La Palma
obtained in 2008 with the Wide Field Camera (pixel size
0\arcsec.33). Seeing was 1\arcsec.0 FWHM.  The image of A~63 was
obtained in 1995 at the 3.5m ESO/NTT with EMMI (exposure time 30~min,
pixel size 0\arcsec.27, seeing 0\arcsec.9). The image of Ou5 is the
one described in \citet{c14}. The inner regions of the nebulae are
displayed in Figure~\ref{F-neb}.

Spectra of A~46, A~63 and Ou5 were obtained on August 16--18 2014 with
the 4.2m WHT telescope and the double-arm ISIS spectrograph. The long
slit of ISIS was opened to 1\arcsec\ width and positioned through the
brightest parts of the nebulae as indicated in Figure~\ref{F-neb} and in
Table~\ref{T-4neb}.  In the blue arm of ISIS, grating R1200B was used,
providing a dispersion of 0.22~\AA~pix$^{-1}$ and a resolution of
0.8~\AA. In the red arm, grating R316R gave a dispersion of
0.92~\AA~pix$^{-1}$, and a resolution of 3.4~\AA.  Two different
grating tilts were adopted in different nights in order to have a
global wavelength coverage from 3610 to 5050~\AA\ in the blue, and
from 5400 to 9150~\AA\ in the red, with little vignetting at the
wavelength ends.  The spatial scale was $0\farcs2$~pix$^{-1}$ in the
blue, and $0\farcs22$~pix$^{-1}$ in the red.
Depending on the wavelength setting, total exposure times were of 2
hours for A~46, 2 to 3.3 hours for A~63, and 2 to 4.7 hours for Ou5.
The spectrophotometric standards BD+28~4211, BD+33~2642, and Feige 110
from \citet{oke90}, and HR~718 from \citet{h94} were observed for flux
calibration.  Seeing varied significantly during observations, and
only data obtained with a seeing value $\le$$1\farcs6$ FWHM were
retained.  Images and spectra were reduced using packages {\it ccdred}
and {\it twodspec} in IRAF\footnote{IRAF is distributed by the
  National Optical Astronomy Observatory, which is operated by the
  Association of Universities for Research in Astronomy (AURA) under
  cooperative agreement with the National Science Foundation.}.

\begin{figure*}[!ht]
\epsscale{2.18}
\plotone{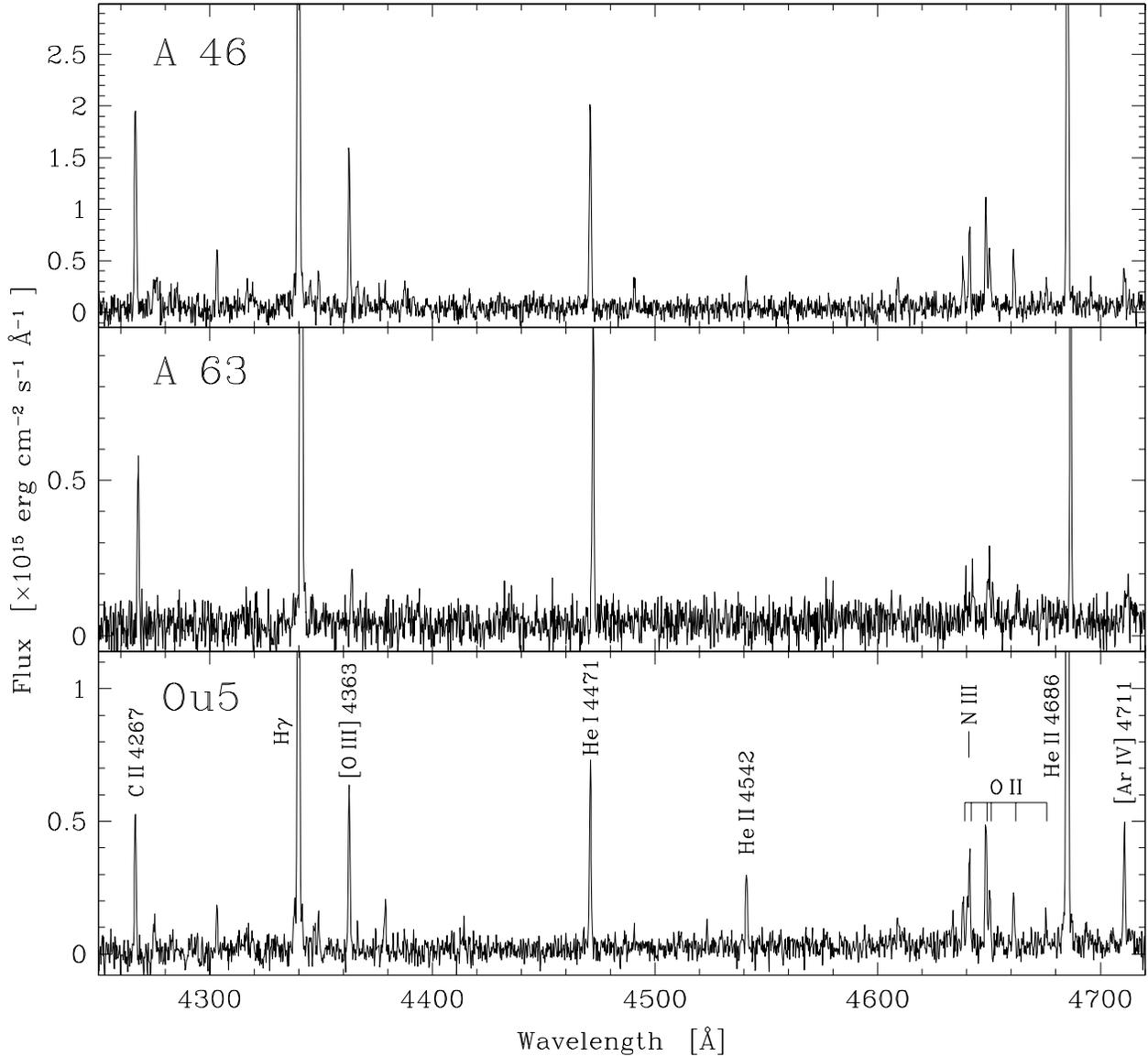}
\caption{Illustrative portion of the spectra of the nebulae integrated
  along the slit.}
\label{F-spectra}
\end{figure*}

\section{Data analysis}\label{S-anal}

To increase the signal from these low surface brightness
nebulae, 1-D spectra were extracted by integrating all the nebular
emission included in the long slit. Note that in A~46, at the position
indicated by the arrow in Figure~\ref{F-neb}, the slit intersects an
elongated knot. In the spectra, this feature clearly stands out for
its strong {\fnii} 6548,6583, {\fsii} 6716,6731, and more moderate
{\foii} 3726,3729 emission relative to the surrounding nebula, from
which it seems to be blue-shifted by $\approx$10~\kms. It could be a
low-ionization small-scale structure (LIS) as often found in PNe
\citep{gcm01}. Other similar blobs are seen in the image, but it
should also be noted that the field is also rich in small background
galaxies.  This feature does not affect the following analysis, and is
not considered any further in this work.

Figure~\ref{F-spectra} shows an illustrative portion of the spectra of
the nebulae, in a wavelength range where some of the most relevant
ORLs and CELs are located.  Emission-line fluxes were measured by
means of multi-Gaussian fit using {\it splot}. Both in the blue and
red, fluxes measured in the two spectrograph settings were
rescaled using the emission lines in the
overlapping spectral range.  The flux differences in these overlapping
lines (after scaling) were used to estimate the errors in the flux
measurement. Identification of the observed lines, their fluxes and
errors are presented in Table~\ref{T-fluxes}.  All line fluxes are
normalised to \hb=100, where the observed \hb\ flux integrated along the
slit is 3.53$\times$10$^{-14}$, 1.80$\times$10$^{-14}$, and
1.37$\times$10$^{-14}$ erg~cm$^{-2}$~s$^{-1}$ for A~46, A~63, and Ou5,
respectively.

The nebular reddening was computed from the hydrogen line ratios,
specifically the Balmer \ha, \hb, \hg, and \hd\ lines and the Paschen
10--3, 11-3 and 12--3 transitions. The reddening law of \citet{f04}
was adopted.  The theoretical hydrogen line ratios have a small but
non-negligible dependence on the electron density and temperature
\elecd\ and \te.  As discussed below, in addition to a standard
nebular component at \te$\sim$10$^4$~K, in these nebulae there is
evidence for a much colder component at $\approx$10$^3$~K, as for
instance indicated by the Balmer jump in A~46
(Sect.~\ref{S-physchem}).  We therefore computed the reddening by
adopting either \te=12500~K, characteristic of the CELs, or
\te=1000~K, as suggested by the ORLs indicators.  The mean values of
the logarithmic extinction at \hb\ of each nebula (\cb) are listed in
Table~\ref{T-phys1} for both temperature assumptions. Those derived
using the CELs temperature are 0.17~dex larger, and in good agreement
with the previous estimates by \citet{pb97} and \citet{c14}, as
expected considering that they were determined under the same
assumption. In the case of A~46, they are also consistent with
  the reddening maps of \citet{1998ApJ...500..525S}, 
which indicate a total foreground dust column corresponding to \cb=0.17.
In the following, we conservatively use the \cb\ values determined
with the CELs temperature to deredden the observed line fluxes, as the
ORLs temperature are more uncertain, especially for A~63 and
Ou5. However, we also repeated the whole analysis with the lower
\cb\ values resulting from the ORLs temperature. Differences
in the results are small and do not affect our conclusions.
Errors on the dereddened fluxes
includes the contribution of the uncertainty in the \cb\ value.

\section{Physical and chemical properties of the nebulae}\label{S-abund}
\label{S-physchem}

\begin{figure}[!ht]
\epsscale{1.0}
\plotone{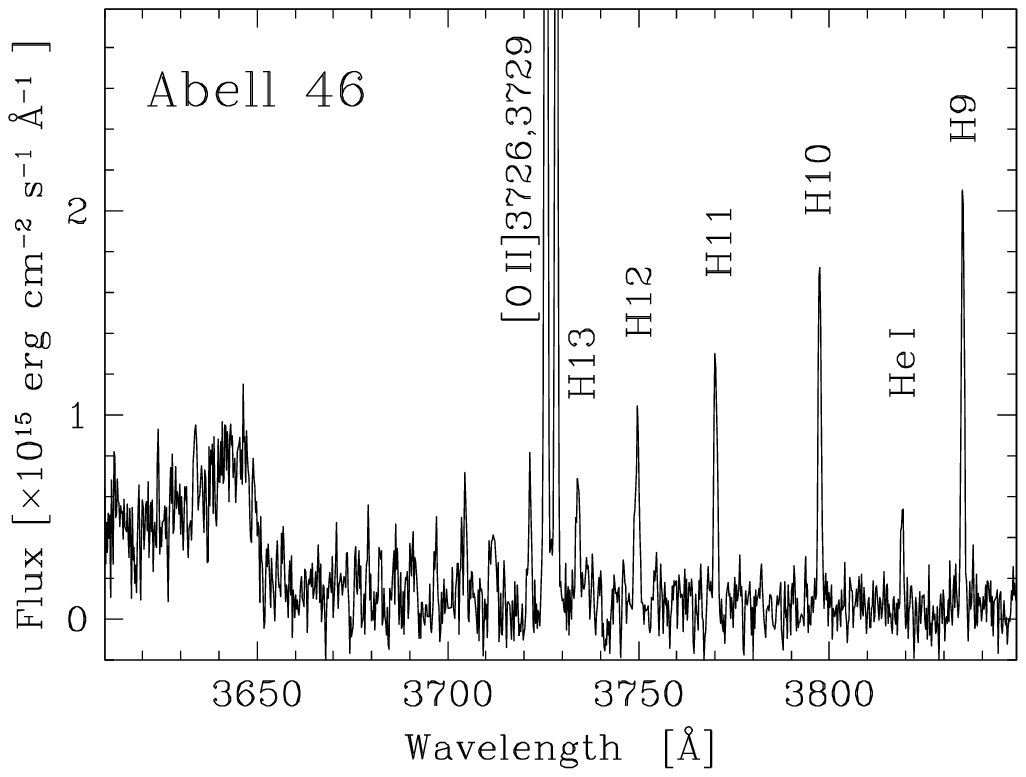}
\caption{Spectrum of A~46 at the Balmer jump.}
\label{F-a46bal}
\end{figure}

\subsection{Physical conditions}

The physical conditions for our nebulae were computed using several
emission-line ratios.  The electron temperatures, {\te}, and electron
densities, {\elecd}, are presented in Table~\ref{T-phys1}.  The
computations of physical conditions were done with PyNeb
\citep{luridianaetal14}, a python based package that allows derivation
of the physical conditions and ionic and elemental abundances from
emission lines.  The methodology followed for the derivation of {\elecd}
and {\te} has been described in \citet{garciarojasetal12}. We used the
state-of-the-art atomic data listed in Table~\ref{atomic_cels}. Errors
in these diagnostics were computed via Monte Carlo simulations.

We computed {\elecd}({\foii}) and {\elecd}({\fsii}) for the three
nebulae, and with lower precision {\elecd}({\fariv}) for A~46 and
Ou5. Similarly to what found by \citet{l06} for Hf~2--2, we find that
{\elecd}({\foii}) is significantly higher than {\elecd}({\fsii}) and
{\elecd}({\fariv}).  For the following analysis, we adopted the
weighted average of {\elecd}({\foii}) and {\elecd}({\fsii}) for A~46
and A~63, and of {\elecd}({\foii}), {\elecd}({\fsii}) and
{\elecd}({\fariv}) for Ou5. Note that the main results are almost
independent on the assumed {\elecd}.


The intensity of the auroral {\fnii} $\lambda$5755 line and the
trans-auroral {\foii} $\lambda$$\lambda$7320+30 lines were corrected
for the contribution of recombination using equations (1) and (2) in
\citet{liuetal00}.  A preliminary calculation of the N$^{2+}$/H$^+$
and O$^{2+}$/H$^+$ abundance ratios by means of {\nii} multiplet 3 and
{\oii} multiplet 1 was used to compute the recombination contribution
to {\fnii}5755 and {\foii} 7320+30, respectively. This resulted
(Table~\ref{T-recombcontrib}) in relatively small corrections for the
{\fnii} line flux, but very large corrections for the {\foii} line
fluxes. This means that \te\ from the corresponding temperature
diagnostics would be lower than computed neglecting recombination.  In
particular, the fact that recombination dominates the trans-auroral
{\foii} emission (100\%\ in A~46 and Ou5, and 56\%\ in A~63), no
reliable \te\ estimation can be derived using these lines.
{\te}({\fnii}) is also very uncertain owing to the large errors in the
measurement of the faint {\fnii}\,5755 and multiplet 3 {\nii} lines.

Given the relatively high ionization degree of our three objects, we also
computed the contribution of recombination to the auroral
\foiii$\lambda$4363 line using equation (3) in \citet{liuetal00}. To
estimate the O$^{3+}$/H$^+$ ratio, we assumed
\begin{displaymath}
\mbox{O}^{3+}/\mbox{H}^+=(\mbox{He}/\mbox{He}^+)^{2/3}\times(\mbox{O}^+/\mbox{H}^++\mbox{O}^{2+}/\mbox{H}^+).
\end{displaymath}
Using He$^+$/H$^+$, He$^{2+}$/H$^+$, and O$^{2+}$ /H$^+$ ratios from
ORLs (see Table~\ref{T-ionic}), and O$^+$/H$^+$ from ORLs in Ou5 and
from CELs rescaled to ORLs using O$^+$/O$^{2+}$ CELs ratio for A~46
and A~63 (see Table~\ref{T-ionic}), the contribution to
\foiii$\lambda$4363 is estimated to be between $\sim$14\%\ and
$\sim$43\%\ depending on the nebula for the slit-integrated spectra
(Table~\ref{T-recombcontrib}, but see also discussion in
Section~\ref{S-profiles}). The values obtained for {\te} with and
without recombination correction are summarized in Table~\ref{T-phys1}.
Owing to the large uncertainties in {\te}({\fnii}) and {\te}({\foii}),
and considering the overall high excitation of the nebulae, we finally
adopted {\te}({\foiii}) as representative of the whole nebula, (see
Table~\ref{T-phys1} and discussion below).

To test whether in these nebulae part of the ORLs emission
comes from a cold, high-metallicity gas component \citep[e.g.][]{l06},
we also checked some {\te} diagnostics from ORLs. Unfortunately,
the most widely used diagnostics, that is the ratio between the
Balmer discontinuity at 3646~\AA\ and the Balmer decrement ratio, could 
only be measured in A~46 (Figure~\ref{F-a46bal}).  In this nebula, the
Balmer continuum temperature (in K) was determined following the
equation by \citet{liuetal01}:
\begin{displaymath}
T({\rm Bac}) = 368 \times(1 + 0.259y^+ + 3.409y^{2+})\left( \frac{{\rm Bac}}{{\rm 
H11}}\right)^{-3/2}
\end{displaymath}
where $y^+$ and $y^{2+}$ are the He$^+$/H$^+$ and He$^{2+}$/H$^+$
ratios respectively, and $Bac$/H11 is the ratio of the discontinuity
of the Balmer jump in erg cm$^{-2}$ s$^{-1}$ \AA$^{-1}$ to the {\hi}
H11 line flux. For A~46, a very low {\te}(Bac) = 1150\,$\pm$550~K, is
obtained.
We also considered other {\te} line diagnostics from
recombination lines, such as {\te} derived from {\hei} line ratios
\citep[{\te}({\hei} 5876/4471, 6678/4471),][]{zhangetal05, l06} and
      {\te} derived from the ratio of {\oii} ORLs to {\foiii} CELs
      \citep[{\te}({\oii} V1/F1),][]{peimbertetal14}.  Additionally,
      for A~46 we could measure the {\te} sensitive ratio {\oii}
      4089/4649 \citep{w03}.
As it can be seen in Table~\ref{T-phys1}, all these diagnostics give
values of the electron temperature that are much lower than those
derived from CELs diagnostics. All this clearly points to the presence
of a nebular component of cold ionized gas where most of the ORL
emission arise, as it has been claimed by e.g. \citet{l06}.

\subsection{Chemical abundances}

Ionic chemical abundances from CELs were computed using PyNeb
\citep{luridianaetal14} and the atomic data in
Table~\ref{atomic_cels}. Errors in the line fluxes and the physical
conditions were propagated via Monte Carlo simulations.  The average
{\elecd} and {\te}({\foiii}) for the whole nebula computed in the
previous section were used.  Assuming other density values (e.g. the
one measured for a specific ion) would cause very small changes on the
majority of ions, with the exception of O$^+$, S$^+$, Cl$^{2+}$ and
Ar$^{3+}$, whose abundances have a more marked dependence on
\elecd. However, even so the main conclusions of this work would not
change. Ionic abundances are presented in Table~\ref{T-ionic}.

As with recombination lines, He$^+$ and He$^{2+}$ abundances were
computed from the {\hei} 4471, 5876, 6678 \AA\ and {\heii} 4684
\AA\ lines using the updated atomic data in Table~\ref{atomic_rls}.  In
our spectra, we detected and measured several heavy element ORLs,
mainly of {\oii} and {\cii}, but also of {\oi}.  To compute
O$^{2+}$/H$^+$ ratio we used multiplet 1 {\oii} ORLs around 4650~\AA,
which are the brightest and widely used {\oii} ORLs \citep[see
  e.~g.][and references therein]{garciarojasetal13}.  We also computed
the C$^{2+}$/H$^+$ ratio using {\cii} at 4267~\AA; this is a $3d-4f$
transition that is, in principle, purely excited by recombination.
The sources of the recombination coefficients that we have used to
compute the ionic abundances of C$^{2+}$ and O$^{2+}$ are listed in
Table~\ref{atomic_rls}. \citet{tsamisetal03} and \citet{ruizetal03}
pointed out that the upper levels of the transitions of multiplet 1 of
{\oii} are not in local thermodynamic equilibrium (LTE) for densities
$n_e$$<$10$^4$ cm$^{-3}$, which is the case for our three objects, and
the abundances derived from each individual line may differ by factors
as large as four. To recover the correct abundances for individual lines,
we applied the non-LTE corrections estimated by
\citet{apeimbertetal05} to our data.  We also computed the
O$^{2+}$/H$^+$ ratio from the sum of all lines of the multiplet
following the recipe given by \citet{estebanetal98}. This method takes
into account the intensity of the whole multiplet and is not affected
by non-LTE effects. This is the value that we finally adopt as
representative of the O$^{2+}$/H$^+$ ORLs abundance ratio 
(Table~\ref{T-ionic}).
  
Additionally, we computed the {\elecd} that minimizes the dispersion
of individual abundances obtained from the {\oii} multiplet 1
lines. This quantity, labelled as {\elecd}({\oii}), is shown
in Table~\ref{T-phys1}. For A~46 and A~63, {\elecd}({\oii}) is very
similar to {\elecd}({\foii}). In Ou5, {\elecd}({\oii}) is
higher than electron densities derived from CELs, in agreement with
what found in Hf~2--2 by \citet{l06}.

Finally, in Ou5 we also computed the O$^+$/H$^+$ ORLs ratio (see
Table~\ref{T-ionic}), as we detected an emission line that we identify
as the blend of three {\oi} ORLs at $\sim$7772 \AA\ (see
Tab~\ref{T-fluxes}).  These lines are hardly detected in deep emission
spectra of other PNe and {\hii} regions, mainly because they are
intrinsically faint and frequently blended with strong
telluric emission lines (which is not the case in the spectrum of Ou5). 

The O$^{2+}$ \adf\ was computed for each object from the comparison of
the ORLs and CELs ionic abundances.  In Ou5, we could also compute the
\adf\ for O$^+$.  Results are shown in Table~\ref{T-ionic}.  It is
clear that the \adfs\ of these nebulae-- and especially in A~46 and
Ou5-- are very large, independently of adopting or not a correction
for recombination in the calculation of {\te}.
A detailed discussion on the possible origins of the large \adfs\ of
these objects, the largest ones found in PNe and two orders of
magnitude larger than in \hii\ regions, is presented in
Sect.~\ref{S-discuss}.

\subsubsection{Total abundances}

Total abundances for N, C, O, Ne, S, Ar and Cl were computed using
the ionization corrections factors (ICFs) published by
\citet{di14} from a large grid of photoionization models of PNe. 
The He abundance was computed by simply adding the He$^+$ and
He$^{2+}$ abundances.  Total abundances are presented in
Table~\ref{T-totab}.

\subsection{Spatial variations}
\label{S-profiles}

\begin{figure}[!t]
\epsscale{1.0}
\plotone{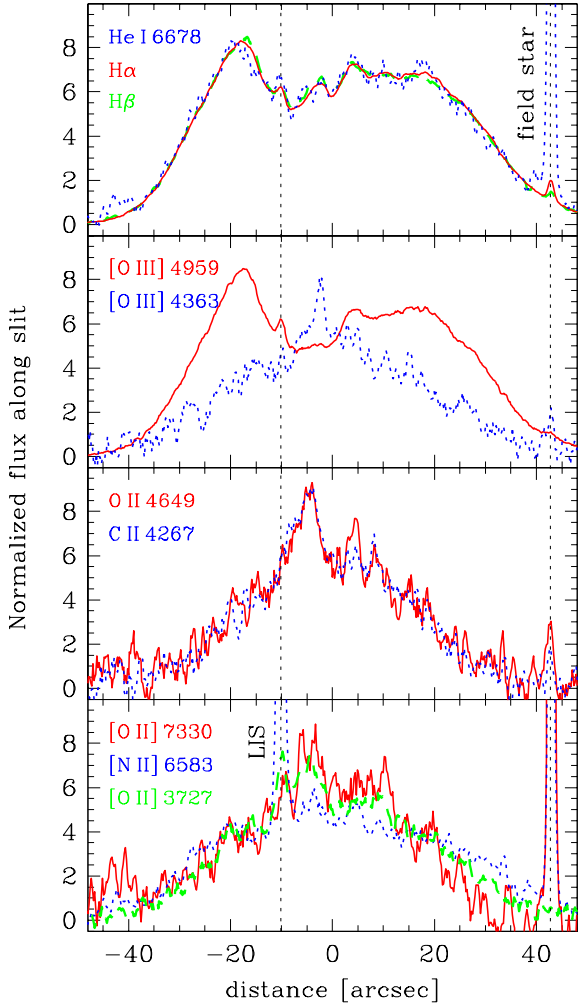}
\caption{Spatial profiles along the slit of selected emission lines
  for A~46. The $x$-axis zero point corresponds to the 
  part of the slit closest to the central star. The emission peak for
  the ORLs {\cii} and {\oii} lines is marked in Figure~\ref{F-neb}.}
\label{F-a46prof}
\end{figure}

\begin{figure}[!t]
\epsscale{1.0}
\plotone{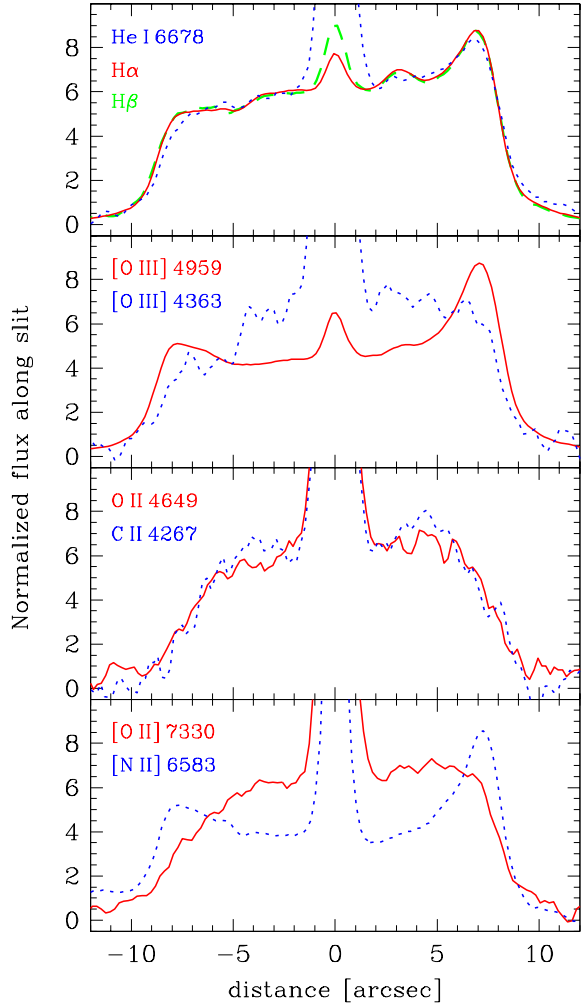}
\caption{Same as in Figure~\ref{F-a46prof}, but for Ou5. 
The central peak is the continuum emission from the central star.  The
{\foii} 3727 profile is not shown here because of the significant,
asymmetrical contamination of the background Galactic emission.}
\label{F-ou5prof}
\end{figure}

The low surface brightness of these nebulae does not allow us to study
in detail spatial variations of the chemical abundances and \adfs.
Useful insights can however be gained by inspecting the spatial
distribution of selected CELs and ORLs along the spectrograph slit.
They are presented in Figs.~\ref{F-a46prof} and \ref{F-ou5prof} for
A~46 and Ou5. To decrease the noise in the faintest emission lines,
profiles were smoothed with a boxcar average of 5 pixels (1\arcsec)
width for A~46, and 3 pixels (0\arcsec.6) for Ou5. Profiles have been
arbitrarily scaled in flux to plot all lines using the same display.
A~63 is not shown because of the low S/N of several of these lines.

In the light of \ha\ (Figure~\ref{F-neb}), A~46 looks amorphous and
diffuse. Even so, its brightest regions seem to trace the outline of a
low surface brightness inner ``cavity'' with a bipolar shape whose
long axis is approximately oriented at position angle P.A.=45\degr.
Figure~\ref{F-a46prof} shows that \hi, \hei\ and strong CELs such as
{\foiii} 4959 have a similar spatial distribution that follows what is
seen in Figure~\ref{F-neb}.  On the other hand, the ORLs emission,
illustrated by {\cii} 4267 and {\oii} 4649, peaks in the innermost regions,
inside the low surface brightness \ha\ ``cavity''. Their peak is 
indicated by a green cross in Figure~\ref{F-neb}. This central
enhancement is a standard characteristic of the ORLs abundances
\citep{l03,t08}.  Most of the fluctuations observed in the ORLs spatial
profiles are noise in these faint lines, which prevents
the detection of small-scale spatial variations such as for instance
done by \citet{t08} in NGC~7009.  More puzzling is the behaviour of
the auroral line {\foiii} 4363, which shows a very similar spatial
distribution to the ORLs, even if its peak is displaced to
the west by a couple of arcseconds.  The marked difference between the
{\foiii} 4959 and {\foiii} 4363 spatial profiles apparently indicates that
\te(\foiii) decreases from 20\,000~K in the innermost regions to
11\,000~K in the outer parts.  However, in the previous section we
have shown that auroral lines can be significantly affected by
recombination. Table~\ref{T-recombcontrib} provides some crude
estimates for the integrated spectrum, but the actual contribution of
recombination to each line as a function of position can only be roughly 
estimated with the present data. The spatial profile of {\foiii} 7330,
whose emission should be dominated by recombination, is also very
similar to that of the ORLs and qualitatively supports the hypothesis
that at least part of the apparent \te(\foiii) radial gradient could
be spuriously produced by variable contribution of recombination to
the auroral lines.

This is confirmed by the following exercise. We extracted the spectrum
within 7 arcsec of the ORLs peak in the spatial profiles of A~46. Even
if some line fluxes are one order of magnitude smaller than in the
integrated spectra, most of the critical lines could be measured. By
proceeding as for the integrated spectra, 48\%\ of the \foiii4363 flux
is estimated to be produced by recombination and was corrected
accordingly.  The value \te(\foiii)=17800$\pm$1900~K that would be
obtained by neglecting recombination is then lowered to
12950$\pm$1150~K when it is instead taken into account.  A
\te(\oii~V1/F1) of 3840$\pm$200~K is computed, which is only slightly
lower than for the integrated spectrum. Densities are about twice as
high, namely \elecd(\fsii)= 560$^{+2200}_{-450}$ cm$^{-3}$ and
\elecd(\foii)=5750$^{+6000}_{-2950}$ cm$^{-3}$.  Total
abundances determined from CELs only vary slightly compared to those
found using the integrated spectrum. But, as far as the O$^{2+}$
abundance is concerned, the CELs abundance slightly decreases to
7.70$\pm$0.11 while the ORL one more markedly increases to
10.21$^{+0.17}_{-0.19}$. This results in an \adf\,(O$^{2+}$) over 300
in the inner regions of A~46!

Ou5, which has a clean bipolar geometry \citep{c14}, shows
similar results (Figure~\ref{F-ou5prof}). \hi, \hei, \foiii4959, and
\fnii6583 have a similar spatial distribution, peaking at the
limb-brightened edges of the bipolar lobes.  On the other hand, ORLs
such as \cii\ and \oii, as well the auroral lines of \foiii\ and
\foii\ are more concentrated toward the center, and are mutually
similar. Therefore recombination should be carefully
considered when the auroral lines are used to determine the
physico-chemical conditions in this kind of nebulae.

\section{Total \ha\ flux and nebular mass}
\label{S-mass}

A~63 and Ou5 were observed by the IPHAS \ha\ photometric survey of the
Galactic plane \citep{d05}. Therefore their IPHAS \ha+\fnii\ images
can be flux calibrated using the information in the IPHAS Second Data
Release \citep{b14}.  The observed, total \ha\ fluxes of these
nebulae, after removing the contribution of {\fnii}\,6548,6583 to the
IPHAS \ha\ filter using our spectroscopy, are listed in
Table~\ref{T-4neb}. Errors are of the order of 10\%.
Taking advantage of the fact that the image of A~46 in
Figure~\ref{F-neb} was obtained with the same instrumentation and filter
in a clear night, we adopted the mean value of the zeropoint of the
IPHAS survey to also calibrate this image. The additional error
introduced by this assumption is another 10\%.

With the total \ha\ flux and the physical parameters determined from
spectroscopy, and adopting the distances quoted in Table~\ref{T-4neb},
the nebular ionized mass can be estimated.  The total
ionized mass of hydrogen was computed by applying the expression:
\begin{displaymath}
m_{neb}(H^+) = \frac{4 \pi\, D^2\,\, F({\rm H\beta})\,\, m_p}{h \nu_{\rm H\beta}\,\, n_e \,\,{\alpha}^{eff}_{\rm H\beta}(H^0, T_e)}
\end{displaymath}
where $F$(H$\beta$) is the dereddened H$\beta$ flux, D
is the distance to the object,
$m_p$ is the proton mass, {\elecd} is the electron density, $h\nu_{\rm
  H\beta}$ is the energy of an H$\beta$ photon and, finally,
$\alpha_{\rm H\beta}^{eff}$($H^0$, {\te}) is the effective
recombination coefficient of H$\beta$.  The usual dependence on the
square of the distance appears in the formula. In this respect, it
should be noted that the distances of A~46 and A~63 were originally
computed through the stellar parameters of the central binaries, which
should provide more reliable determinations than any statistical
method, especially when the peculiar nature of these nebulae is
considered.

As $\alpha^{eff}_{\rm H\beta}(H^0, T_e)\propto$\,\te$^{\sim-1}$,
the total mass is roughly proportional to the adopted
electron temperature \te\ and inversely proportional to the electron
density \elecd.  The dual-component nature of these nebulae prevents a
precise mass determination, as the relative amounts of \hi\ emitting
at the low \te\ indicated by the Balmer jump and the \hei\ or
\oii\ ORLs, and that emitting at the much higher CELs \te\ is not
known. Strictly speaking, at least for A~46 one would be tempted to
consistently adopt for hydrogen the temperature computed using the
Balmer jump, namely 1000~K. As with densities, there is also a clear
difference (by up to a factor of 8) between those computed from
\foii\ and those computed from \fsii. 
Assuming \te=1000~K and the average \elecd\ 
from CELs, we determine an H$^+$ mass of 
1$\times$10$^{-3}$, 
5$\times$10$^{-4}$, and 
5$\times$10$^{-3}$~\msun\ for A~46, A~63, and Ou5, respectively. 

However, even if the total amount of metals in the hot and cold gas
phases is comparable as for instance estimated for Hf~2--2
\citep{l06}, the much higher metallicity of the cold component implies
that the majority of hydrogen must be in the hot component. It seems
therefore reasonable to adopt \te(\foiii), which would imply H$^+$
masses almost one order of magnitude larger, namely
8$\times$10$^{-3}$, 4$\times$10$^{-3}$, and 5$\times$10$^{-2}$
~\msun\ for A~46, A~63, and Ou5, respectively.

Finally, in the less likely case that all hydrogen comes from a plasma
with \te(\foiii) and the lowest electron density measured (\elecd(\fsii)), 
we obtain 
4$\times$10$^{-2}$,
1$\times$10$^{-2}$, and 
2$\times$10$^{-1}$ ~\msun, respectively. 
These should be just taken as conservative upper limits to the
total amount of H$^+$ in these nebulae (for the adopted distances).
\citet{l06} found a similarly low H$^+$ mass of
1.4$\times$10$^{-2}$~\msun\ for Hf~2--2. 

Therefore, in spite of the large uncertainties in these calculations,
the clear conclusion is that these nebulae with large \adfs\ and close
binary central stars have a very small total
ionized mass, one or more orders of
magnitudes lower than typical of Galactic PNe \citep[see
  e.g.][]{pot84}.  The total nebular mass could in principle be larger
if part of it is in neutral/molecular form. However, spectral
indicators such as the absence of \foi\,6300 emission
\citep{hs66,kj89}, which is thought to be produced at the interface
between ionized and neutral gas via charge exchange reactions, point
to optically thin, density bounded nebulae.

\section{Discussion}\label{S-discuss}

Two of the three PNe that we studied, namely A~46 and Ou5, have
abundance discrepancy factors that are among the largest ones found in
PNe. In particular, A~46 has the largest \adf\ ever measured, with the
exception of the hydrogen-deficient knots that are directly observed
in Abell 30 \citep{w03}.  Our third target, A~63, has a lower
\adf\ ($\sim$10), but still above the average value in PNe.

Our three target PNe have close binary central stars with orbital
periods around ten hours. Therefore these observations, together with
the additional case of Hf~2--2 \citep{l06}, reveal a clear connection
between the most extreme representatives of the abundance discrepancy
problem and the binarity of the central sources. The abovementioned
Abell 30 (and the morphologically similar Abell~58 aka V605~Aquilae,
which experienced a nova-like outburst in 1905) were also proposed to
contain a close binary \citep{w08}, but no direct evidence has been
found yet.

Our spectroscopic analysis supports the previous interpretation that
two different gas phases coexist in these nebulae
\citep[e.g.][]{l06,t08}: hot gas at 10$^4$~K with standard metallicity
where the CELs can be efficiently excited, and a much cooler
(10$^3$~K) plasma with a highly enhanced content of heavy element
abundances (which cause the cooling) where only ORLs form. How much
each component contributes to the total mass, and how they are
distributed and mixed, is poorly known, because existing data are not
able to spatially resolve them. One common characteristic, also shared
by the nebulae studied in this work, is that the ORLs emission is
enhanced in the central regions of the nebulae, where the
\adf\ reaches a value over 300 in A~46. \citet{l06} estimated that
similar amounts of metals are in the hot and cold gas of Hf~2--2, but
the amount of H and He in each component cannot be determined.  This
prevents the calculation of precise X$^{i+}$/H$^+$ abundances for each
gas phase individually \citep{l06}, and therefore all values in
Table~\ref{T-totab} should be taken with caution.  For instance, the
large helium content that is measured in A~63 and Ou5 may not be
completely real \citep[see also][]{l03,t08}. Only ionic abundance
ratios, such as for example N$^+$/O$^+$, are free of this bias. In
this respect A~46, and to a lesser extent A~63 (because of its large
errors) and Ou5\footnote{The much lower N$^+$/O$^+$ ratio for this
  nebula computed by \citet{c14} was caused by the use of the auroral
  \foii7320,7730 lines, that we have shown here to be severely
  affected by recombination processes.}, have low
N$^+$/O$^+$\,$\approx$\,N/O CELs abundance ratios, typical of
low-metallicity SMC PNe, or alternately indicating substantial oxygen
enhancement compared to nitrogen. But even this ratio should be taken
with caution, considering the high excitation of the nebulae.  We have
also noticed that CELs temperature gradients within a nebula can be at
least partly spurious, caused by a varying contribution of
recombination to the emission of the auroral lines. This additional
uncertainty also affects the determination of the CELs abundances.

\subsection{The nature of these nebulae and the implications 
for the abundance discrepancy problem}
\label{S-discuss1}

The main result of this work is that any explanation of the abundance
discrepancy problem must include the fact that {\it the most
  pathological cases are planetary nebulae with a close binary central
  star.}  This casts light on the origin of the problem. Let's
consider previously proposed and new scenarios, and their {\it pro et
  contra}.

A key point to understand the evolution of these systems is the small
ionized mass of the nebulae (Sect.~\ref{S-mass}).  Let's take
10$^{-2}$~\msun\ as a representative figure of the nebular mass for
the systems that we have considered. The total mass in the
high-metallicity component would then be a (likely small) fraction of
it.  These low masses dispute the usual interpretation of the origin
of these nebulae. It is generally thought that PNe with close binary
central stars went through a common-envelope phase when the
radius of the primary star increased in the AGB phase and engulfed the
companion.  The CE interaction caused the shrinkage of the orbit to
the current small separations ($\sim$2~\rsun\ for the four objects in
Table~\ref{T-4neb}) and provoked the complete ejection of the envelope,
forming the PN.  However, the nebular masses that we have estimated
are far too low compared to the AGB envelope masses expected for
0.5-0.6~\msun\ cores and standard initial-to-final mass relationships,
even if some mass loss prior to the AGB is allowed \citep{dm11}.  The
exercise has been done by \citet{a08} and \citet{dm11} for A~46 and
A~63, for which stellar parameters are known (Table~\ref{T-4neb}).  In
\citet{dm11}, consistency between the stellar evolution and binary
parameters is reached for CE masses of 0.4~\msun\ for A~46 and
1.2~\msun\ for A~63, i.e.  one to three orders of magnitude larger,
respectively, than what we measured.
The similar or only slightly larger nebular masses of Ou5 and Hf~2--2
add strength to this conclusion, even if the stellar parameters for
these two systems are not precisely known\footnote{For Ou5, they are
  very likely in the range of A~46 and A~63 (Corradi et al. 2015, in
    preparation).}. This missing-mass problem is even more severe
considering that the whole CE is thought to be ejected in a very short
timescale \citep[$\sim$few years, see e.g.][]{sand98}, limiting the
possibility that the PN is dispersed over a large volume.


As it is evident that the systems at some point went through a CE
phase (given the present small orbital separation), the envelope of
the primary star would have been ejected a long time ago and would have
now vanished in the interstellar medium. Note that the
$0.5$~\msun\ mass of the central star of A~46 is at the limit between
RGB and AGB core masses. It cannot be excluded that in this case (but
can be ruled out for A~63 which has a typical AGB core mass of
0.63~\msun) the CE interaction, ejection, and orbital shrinking
occurred in the RGB phase of the primary.

Then the currently observed low--mass nebulae would be produced much
after the CE phase, in post--CE mass loss episodes
when the massive AGB envelope has already
dispersed. Even so, their central stars have to be still hot and
luminous to be able to ionize the nebulae, as it is indeed observed
\citep[see e.g.][]{a08}.


\subsubsection{Very late thermal pulses (VLTPs)}

This hypothesis has been discussed in several articles
\citep{l06,w08}.  In general, thermal pulses in the post--AGB phase
are expected to be relatively frequent \citep{b01}. In the best
studied case, Sakurai's object, the estimated ejected mass is
2$\times$10$^{-3}$~\msun\ \citep{h05}, i.e., in the possible range of
what we have measured.  Gas falling back from the ejected envelope,
accreting on the post-AGB star, and perhaps triggering a VLTP was
predicted and discussed by, e.g., \citet{fs09}. In those specific
calculations, however, the total amount of fallback material,
10$^{-5}$ to 10$^{-3}$~\msun, seems on the low side of what we
observed, and the timescale for accretion is too short to allow the
massive, ejected CE to vanish completely.

The VLTP ejecta are expected to basically represent the intershell
abundances, with C/O$\gg$1. In the case of Ou5, the ORL C/O ratio
  could be estimated as C/O=$a$\,(C$^{2+}$/O$^{2+}$), where $a$=1.12
  is the ICF from \citet{di14} that could be consistently computed
  using the ORL O$^{+}$ and O$^{2+}$ abundances.  We obtain
  C/O=0.55$\pm$0.24 for this nebula.  For A~46 and A~63, no ORL
  O$^{+}$ abundance determination is available, and therefore the ICF
  can only be estimated using CELs which would indicate values very
  close to unity. We obtain C/O=0.78$\pm$0.11 for A~46, and
  1.66$\pm$0.47 for A~63 which is the only nebula where the C/O ratio
  seems to be above unity. Therefore, in spite of the uncertainty
  involved in the use of the faint ORLs, the low C/O ratios estimated
  in A~46 and Ou5, as well as in Hf~2--2 \citep{l06}, argue against
  VLTPs as the origin of the high-metallicity ejecta causing the large
  observed \adfs.

\subsubsection{Nova-like outbursts}

Nova outbursts are natural phenomena in interacting binaries. They
result from thermonuclear runaways on the surfaces of white dwarfs
(WDs) accreting hydrogen-rich matter in binary systems \citep[see
  e.g.][]{g98}.  Nova ejecta are metal enriched by mixing between the
accreted envelope and the WD core. Therefore, in principle, the
metal-rich gas in our target nebulae could be ejected in nova
outbursts from the close binary central stars, as suggested by
\citet{w03} and \citet{w08} for A~30 and A~58. In these two cases, the
hypothesis of a nova eruption could explain the large neon enrichment
of their hydrogen-deficient knots, if produced in a massive
(M$\ge$1.2~\msun) ONeMg WDs. For A~58, this scenario (in combination
with a final helium shell flash) or the alternative possibility of a
stellar merger are evaluated in \citet{ldl11}.

However, there is no indication that mass transfer from the
secondaries to the primaries is currently going on in the systems that
we have studied, because the secondary stars -- albeit inflated -- do
not fill their Roche lobe \citep{a08}.  To feed the WD, accretion must
therefore come from circumbinary material left-over in the systems
from the CE phase.  A possibility is that the envelope is not removed
completely in the CE interaction and part of it remains
gravitationally bounded to the system.  The non-accreted part of this
circumstellar gas, with normal AGB abundances, could then interact
with the nova ejecta and partially mix with it producing the observed
dual-component nebulae.  In this respect, it should be noted that the
existing simulations of the CE process fail to eject the entire
envelope, a large fraction of which remains gravitationally bound to
the system \citep{sand98,pa12,rt12}. This may result in a series of CE
and fall-back events, and in the formation of a left over, low mass
circumbinary disk \citep[][Kuruwita, De Marco \& Staff, in
  preparation]{ks11,ku15}, which could in turn form metal-rich dust as
perhaps observed in the putative post-CE system V1309~Sco \citep{n13}.

Possible problems are the very low mass of classical nova ejecta
\citep[$<$10$^{-3}$~\msun, but generally much smaller, see
  e.g.][]{g98}, and the short timescale (a few years) of their nuclear
burning and high luminosity phase that are needed to light up
the PNe.  Long-lasting outbursts at high WD luminosity can be achieved if
the mass accretion rate is higher than in classical novae, as for
instance occurs in symbiotic (very slow) novae. The relevant
parameters space should be explored in detail to see how this
hypothesis can fit our post--CE PNe \citep[cf. e.g.][]{y05}.

An additional important parameter is the expansion velocity of the
ejecta.  In the PNe that we have studied, the ORL emitting gas phase
does not show the large expansion velocities (few to many hundred
\kms) typical of classical and symbiotic novae, which would be easily
revealed at the resolution of our spectra. While more precise
information is not available yet, a further constraint for the
modeling is that both the hot and cold gas components have expansion
velocities more typical of PNe than of novae.

\subsubsection{Planetary material}

A source of the metal-rich component in the nebulae might be planetary
debris that survived the CE phase.  Indeed, about a quarter of {\it
  single} WDs have considerable amounts of metals in their
photosphere, that are explained by accretion of tidally disrupted
planetesimals/asteroids/comets that survived through all previous
evolutionary phases of the star \citep[see e.g.][]{f14}.

The possible contribution of solid bodies to the metal-rich component
in PNe was first mentioned by \citet{l03} and then investigated by
\citet{hs10}. They conclude that solid bodies could provide the source
of the high-metallicity gas only if they are meter-sized or larger,
their evaporation process extends back to the final AGB evolution,
mixing with the gas component at ``normal'' AGB metallicity is
inefficient, and systems possess a much more massive population of
comets than is found in our Solar system.

Note also that \citet{t11,t13} and \citet{md12} proposed that
photoevaporating protoplanetary disks are the cause of the smaller
\adfs\ found in \hii\ regions.

\subsubsection{Planets destruction}

Another possibility is that these post-CE nebulae are produced by the
destruction of one or more circumbinary Jovian planets.  There is
increasing evidence that post-CE planets exist \citep{par14,z13,bs14}.
It is therefore possible that planets in unstable orbits spiral in,
are tidally disrupted, and collide with the post-AGB star
\citep{bs12}. The hydrogen-rich envelope of the planet would be
stripped first, and then its metallic core. Part of this material
would be accreted by the WD: at high accretion rates, the gas cannot
be processed and inflates a red-giant envelope, producing a new CE
phase and eventually the ejection of a new nebula with the dual nature
(H-rich from the planet envelope and metal-rich from its core) that we
have observed.  The new CE would also favour the formation of highly
bipolar nebulae such as Ou5 and A~63, by forcing the CE ejection
toward the orbital plane, and/or producing jets from one of the two
stars (as observed in A~63).  A possible problem with this
  hypothesis is the apparently enhanced helium abundance of our target
  PNe (Table~\ref{T-totab}), which is not characteristic of
  planets. However, the He abundances in these high-\adf\ objects are
  likely to be overestimated \citep{t08}.

While this scenario should be taken as highly speculative at this
stage, it has several positive aspects such as the fact that the
process for planet collision is long enough ($>$10$^6$ year) to allow
the CE to vanish, the ejection could be split into different phases
(the planet's atmosphere and core), and the observed nebular masses
are in a plausible range corresponding to 1-10 Jupiter masses. In such
case, Sir William Herschel would have been right in coining the name
{\it planetary} nebulae!

\subsection{Other implications and conclusions}

If our hypothesis of an intimate relation between binarity and large
\adfs\ is correct, we predict that many -- if not all -- the PNe where
large \adfs\ have been measured are post--CE binaries.  One already
confirmed case is NGC~6778 which indeed contains a close binary with
P$_{orb}$=0.15 \citep{m11}.  Other primary candidates to search for
binarity are M~1-42, M~2-36, M~3-26, M~3-32, NGC~6153 and NGC~1501
\citep{l03,t04,e04}, as well as the already mentioned Abell 30 and
Abell 58.  Note that our conclusion does not imply the opposite
hypothesis, i.e. that {\it all} post--CE PNe have large \adfs. For
example, NGC~5189 has a close binary central star \citep{ma15} but a
low \adf\ \citep{garciarojasetal13}. Also, no  oxygen or carbon
recombination lines are detected in relatively deep spectra of
the Necklace binary PN \citep{c11}.

If some of the post-CE nebulae are not genuine AGB envelopes, but
rather post-CE mass loss events or post-RGB systems, the total
fraction of close binary PN central stars would be smaller than
presently estimated \citep[$\sim$15\%][]{m09}, and perhaps more
consonant with the fraction of main-sequence binaries
\citep[][but also see \citealt{bof14}]{dm13,dou14}.

Concluding, we have added new ingredients to the discussion of the
nature of PNe with close binary central stars, the CE evolution, and
the abundance discrepancy problem. The main result of this work is the
clear link between binary stellar evolution and the formation of high
\adfs\ in PNe. At the moment, none of the proposed scenarios to
explain this result clearly stands out, but this is likely because
observations provide still limited constraints. To further progress, a
larger sample of objects should be analyzed to highlight the basic
relationships among the numerous parameters involved. In particular,
more precise determinations of crucial quantities such as the nebular
masses and chemical compositions for each (normal and enriched)
nebular components, and the binary parameters (to better reconstruct
the evolution of the systems), are needed.

Finally, even if the binary nature of the central stars seems to be an
essential ingredient to understand the abundance discrepancy problem
in PNe, it is hard to envisage its relevance to other astrophysical
contexts such as \hii\ regions. This may imply that different causes
should be sought to fully understand this long-standing astrophysical
problem, as proposed by \citet{g07}. On the other hand, as
  discussed in the previous sections, some related phenomena may be
  involved, like for instance the possible contribution of
  protoplanetary disks in \hii\ regions and of planetary debris in
  PNe.

\acknowledgments
This work is based on observations obtained with the 4.2m WHT and 2.5m
INT telescopes of the Isaac Newton Group of Telescopes, operating on
the island of La Palma at the Spanish Observatories of the Roque de
Los Muchachos of the Instituto de Astrof\'\i sica de Canarias.  The
WHT spectra of 14 October 2014 were obtained in service time. Also
based on data obtained from the ESO Science Archive Facility.  We are
extremely grateful to Noam Soker and Orsola De Marco for their
significant contribution, in terms of original ideas and proper
criticism, to the discussion in Sect.~\ref{S-discuss1}. We also
  thank Valentina Luridiana for her help with PyNeb, Grazyna
  Stasi\'nska, Christophe Morisset, and C\'esar Esteban for a critical
  reading of the manuscript, and the referee for his/her very useful
  comments. This research has been supported by the Spanish Ministry
of Economy and Competitiveness (MINECO) under grants AYA2012-35330,
AYA2011-22614, and AYA2012-38700. JGR acknowledges support from Severo
Ochoa excellence program (SEV-2011-0187) postdoctoral fellowship. PRG
is supported by a Ram\'on y Cajal fellowship (RYC2010-05762).




{\it Facilities:} 
\facility{ING:Herschel (ISIS)}
\facility{ING:Newton (WFC)}
\facility{ESO:NTT (EMMI)}



%
%
%

\clearpage
\begin{deluxetable}{lccccccccccc}
\rotate
\tabletypesize{\footnotesize}
\tablecolumns{12}
\tablewidth{0pc}
\tablecaption{Basic properties of the PNe, central stars,  and observational details.\label{T-4neb}}

\tablehead{
Name & PN~G & \mc{2}{c}{R.A.\, (J2000)\, Dec} & Slit P.A.  & D         & P$_{orb}$ & $m_{cs}$ & $m_2$ & $i$ & F(\ha) &  $m_{neb}$(H$^+$) \tnm{a} \\
     &      &             &                   & (deg)      & (kpc)     &  (days)  &(\msun)  &(\msun)&(\degr) & (erg cm$^{-2}$ s$^{-1}$) &(\msun)} 
\startdata
Abell 46      & 055.4$+$16.0 & 18 31 18.29 &$+$26 56 12.9 & 100 & 1.7          & 0.47 & 0.51           & 0.15            & 80             & 8.87$\times$10$^{-12}$ & 10$^{-3}$ to 10$^{-2}$      \\
Abell 63      & 053.8$-$03.0 & 19 42 10.20 &$+$17 05 14.4 & 300 & 2.4          & 0.46 & 0.63           & 0.29            & 87             & 2.29$\times$10$^{-12}$ & 10$^{-4}$ to 10$^{-2}$      \\
Ou5           & 086.9$-$03.4 & 21 14 20.03 &$+$43 41 36.0 & 297 & 5            & 0.36 & \mc{2}{c}{$m_2$/$m_{cs}$$\approx$0.4?}  & 90       & 0.91$\times$10$^{-12}$ & 10$^{-3}$ to 10$^{-1}$      \\
Hf~2--2       & 005.1$-$08.9 & 18 32 30.93 &$-$28 43 20.5 &\nd  & 4.25         & 0.40 & \nd            & \nd             & 25--55         & \nd                   & $\approx$10$^{-2}$ 
\enddata
\tablecomments{For the original sources of the orbital periods, see \anchor{http://www.drdjones.net/?q=node/6}{http://www.drdjones.net/?q=node/6}.
Distances, binary and stellar parameters for A~46 are from \citet{b94}, for A~63 from \citet{a08}, and for 
Ou5 from \citet[][2015 in preparation]{c14}.
Hf~2--2 is not directly analized in this paper, but it is shown here because is the other key target for discussion in Sect.~\ref{S-discuss}. 
Its spectroscopic analysis was done by \citet{l06}, the quoted central star inclination is from \citet{s12}.}
\tablenotetext{a}{For the adopted distance (col. 6), and computed as described in Sect.~\ref{S-mass}.}
\end{deluxetable}

\clearpage
\begin{deluxetable}{rlcrrrrrr}
\tabletypesize{\footnotesize}
\tablecolumns{9}
\tablewidth{0pt} 
\tablecaption{Line identifications, observed fluxes and their \%\ errors\label{T-fluxes}}
\tablehead{
 \colhead{} & 
 \colhead{} & 
 \colhead{} & 
 \multicolumn{2}{c}{Abell 46} &
 \multicolumn{2}{c}{Abell 63} &
 \multicolumn{2}{c}{Ou5} \\
\cline{4-5} \cline{6-7} \cline{8-9}  \\
\colhead{$\lambda_0$ (\AA\ )} &
\colhead{Ion} &
\colhead{Mult.} &
\colhead{F($\lambda$)} &
\colhead{Error (\%)} &
\colhead{F($\lambda$)} &
\colhead{Error (\%)} &
\colhead{F($\lambda$)} &
\colhead{Error (\%)}
}
\startdata
 3634.25 &           {\hei} &           28 &   1.21 &	16 &	\nd &  \nd &	\nd &  \nd  \\
 3679.36 &            {\hi} &          H21 &   0.80 &	23 &	\nd &  \nd &	\nd &  \nd  \\
 3682.81 &            {\hi} &          H20 &   0.67 &	26 &	\nd &  \nd &	\nd &  \nd  \\
 3686.83 &            {\hi} &          H19 &   1.03 &	18 &	\nd &  \nd &	\nd &  \nd  \\
 3691.56 &            {\hi} &          H18 &   1.51 &	14 &	\nd &  \nd &	\nd &  \nd  \\
 3697.15 &            {\hi} &          H17 &   0.95 &	20 &	\nd &  \nd &	\nd &  \nd  \\
 3703.86 &            {\hi} &          H16 &   2.56 &	10 &	\nd &  \nd &	\nd &  \nd  \\
 3711.97 &            {\hi} &          H15 &   2.49 &	10 &	\nd &  \nd &	\nd &  \nd  \\
 3721.83 &         {\fsiii} &           2F &   2.02 &	11 &	\nd &  \nd &   2.42 &	31  \\
 3721.93 &            {\hi} &          H14 &	  * &	 * &	\nd &  \nd &	  * &	 *  \\
 3726.03 &          {\foii} &           1F &  23.91 &	 6 &  11.01 &	11 &  26.01 &	 8  \\
 3728.82 &          {\foii} &           1F &  14.82 &	 6 &   6.61 &	12 &  20.68 &	 8  \\
 3734.37 &            {\hi} &          H13 &   2.20 &	10 &   2.97 &	18 &   3.16 &	25  \\
 3750.15 &            {\hi} &          H12 &   3.88 &	 8 &   3.78 &	16 &   3.12 &	25  \\
 3770.63 &            {\hi} &          H11 &   4.35 &	 7 &   3.64 &	16 &   4.17 &	19  \\
 3797.63 &         {\fsiii} &           2F &   5.49 &	 7 &   6.56 &	12 &   5.16 &	16  \\
 3797.90 &            {\hi} &          H10 &	  * &	 * &	  * &	 * &	  * &	 *  \\ 
 3819.61 &           {\hei} &           22 &   1.39 &	14 &	\nd &  \nd &	\nd &  \nd  \\ 
 3835.39 &            {\hi} &           H9 &   6.60 &	 6 &   7.15 &	11 &   7.26 &	12  \\
 3868.75 &        {\fneiii} &           1F &  39.43 &	 5 &  25.37 &	 9 &  74.32 &	 6  \\
 3888.65 &           {\hei} &            2 &  22.66 &	 5 &  22.78 &	 9 &  22.17 &	 7  \\
 3889.05 &            {\hi} &           H8 &	  * &	 * &	  * &	 * &	  * &	 *  \\
 3964.73 &           {\hei} &            5 &   0.91 &	19 &	\nd &  \nd &	\nd &  \nd  \\ 
 3967.46 &        {\fneiii} &           1F &  12.61 &	 5 &   7.33 &	10 &  21.00 &	 7  \\
 3970.07 &            {\hi} &           H7 &  16.69 &	 5 &  17.41 &	 8 &  16.57 &	 7  \\
 4009.26 &           {\hei} &           55 &   0.67 &	25 &	\nd &  \nd &	\nd &  \nd  \\ 
 4026.21 &           {\hei} &           18 &   3.58 &	 7 &   2.86 &	16 &   3.59 &	19  \\
 4068.60 &          {\fsii} &           1F &   4.38 &	 6 &	\nd &  \nd &   4.30 &	16  \\
 4069.62 &           {\oii} &           10 &	\nd &  \nd &   1.69 &	24 &	\nd &  \nd  \\
 4069.89 &           {\oii} &           10 &	\nd &  \nd &	  * &	 * &	\nd &  \nd  \\
 4076.35 &          {\fsii} &           1F &   2.56 &	 8 &	\nd &  \nd &   2.99 &	21  \\
 4078.84 &           {\oii} &           10 &   0.37 &	 : &	\nd &  \nd &	\nd &  \nd  \\ 
 4085.11 &           {\oii} &           10 &   0.67 &	25 &	\nd &  \nd &	\nd &  \nd  \\
 4089.29 &           {\oii} &           48 &   1.16 &	15 &	\nd &  \nd &	\nd &  \nd  \\
 4097.22 &           {\oii} &           20 &   2.69 &	 8 &	\nd &  \nd &   3.20 &	20  \\
 4101.74 &            {\hi} &           H6 &  26.32 &	 4 &  25.87 &	 7 &  26.43 &	 6  \\
 4104.99 &           {\oii} &           20 &   0.65 &	25 &	\nd &  \nd &	\nd &  \nd  \\ 
 4107.09 &           {\oii} &           62 &   0.50 &	32 &	\nd &  \nd &	\nd &  \nd  \\ 
 4119.22 &           {\oii} &           20 &   1.20 &	15 &	\nd &  \nd &	\nd &  \nd  \\ 
 4132.80 &           {\oii} &           19 &   0.65 &	25 &	\nd &  \nd &	\nd &  \nd  \\ 
 4153.30 &           {\oii} &           19 &   1.12 &	16 &	\nd &  \nd &	\nd &  \nd  \\ 
 4267.15 &           {\cii} &            6 &   6.45 &	 4 &   3.99 &	11 &   5.11 &	12  \\
 4275.55 &           {\oii} &           67 &   1.12 &	15 &	\nd &  \nd &   1.52 &	35  \\
 4276.75 &           {\oii} &           67 &   0.99 &	17 &	\nd &  \nd &	\nd &  \nd  \\ 
 4294.92 &           {\oii} &           54 &   0.56 &	28 &	\nd &  \nd &	\nd &  \nd  \\ 
 4303.61 &           {\oii} &           65 &   1.57 &	11 &   0.93 &	37 &   1.60 &	32  \\
 4303.82 &           {\oii} &           53 &	  * &	 * &	  * &	 * &	  * &	 *  \\
 4317.14 &           {\oii} &            2 &   0.56 &	28 &	\nd &  \nd &	\nd &  \nd  \\ 
 4319.63 &           {\oii} &            2 &	\nd &  \nd &	\nd &  \nd &   1.04 &	 :  \\
 4340.47 &            {\hi} &           H5 &  47.34 &	 3 &  48.15 &	 5 &  49.45 &	 4  \\
 4345.56 &           {\oii} &            2 &   0.46 &	33 &	\nd &  \nd &   1.10 &	 :  \\
 4347.41 &           {\oii} &    2D-2D$_0$ &	\nd &  \nd &	\nd &  \nd &   1.22 &	40  \\
 4349.43 &           {\oii} &            2 &   0.93 &	18 &	\nd &  \nd &   1.44 &	35  \\
 4363.21 &         {\foiii} &           2F &   4.88 &	 5 &   1.07 &	32 &   6.33 &	 9  \\
 4366.89 &           {\oii} &            2 &   0.90 &	18 &	\nd &  \nd &	\nd &  \nd  \\ 
 4379.55 &          {\neii} &           60 &	\nd &  \nd &	\nd &  \nd &   1.84 &	27  \\
 4387.93 &           {\hei} &           51 &   0.71 &	22 &	\nd &  \nd &	\nd &  \nd  \\ 
 4414.90 &           {\oii} &            5 &   0.35 &	 : &	\nd &  \nd &	\nd &  \nd  \\ 
 4416.97 &           {\oii} &            5 &   0.44 &	34 &	\nd &  \nd &	\nd &  \nd  \\ 
 4471.47 &           {\hei} &           14 &   6.49 &	 4 &   6.43 &	 7 &   6.27 &	 9  \\
 4491.23 &           {\oii} &          86a &   0.80 &	20 &	\nd &  \nd &	\nd &  \nd  \\ 
 4541.59 &          {\heii} &          4.9 &   0.76 &	21 &	\nd &  \nd &   2.56 &	18  \\
 4609.44 &           {\oii} &          92a &   0.91 &	17 &	\nd &  \nd &   0.91 &	 :  \\
 4638.86 &           {\oii} &            1 &   1.24 &	13 &   0.59 &	 : &   1.51 &	28  \\
 4640.64 &          {\niii} &            2 &	\nd &  \nd &	\nd &  \nd &   1.30 &	32  \\
 4641.81 &           {\oii} &            1 &   2.19 &	 8 &   0.94 &	32 &   2.68 &	17  \\
 4641.85 &          {\niii} &            2 &	\nd &  \nd &	\nd &  \nd &	  * &	 *  \\
 4649.13 &           {\oii} &            1 &   3.14 &	 6 &   0.75 &	 : &   3.81 &	12  \\
 4650.84 &           {\oii} &            1 &   1.81 &	 9 &   0.53 &	 : &   1.80 &	24  \\
 4661.63 &           {\oii} &            1 &   1.51 &	11 &	\nd &  \nd &   1.89 &	23  \\
 4676.24 &           {\oii} &            1 &   0.72 &	21 &	\nd &  \nd &	\nd &  \nd  \\ 
 4685.68 &          {\heii} &          3.4 &  26.41 &	 1 &   6.83 &	 6 &  71.11 &	 1  \\
 4711.37 &         {\fariv} &           1F &   1.25 &	13 &   1.19 &	26 &   3.61 &	12  \\
 4740.17 &         {\fariv} &           1F &   0.93 &	17 &	\nd &  \nd &   2.73 &	16  \\
 4859.32 &          {\heii} &          4.8 &   1.97 &	 8 &	\nd &  \nd &   3.75 &	11  \\
 4861.33 &            {\hi} &           H4 & 100.00 &	 0 & 100.00 &	 1 & 100.00 &	 1  \\
 4921.93 &           {\hei} &           48 &   1.58 &	10 &   1.78 &	17 &   1.46 &	25  \\
 4958.91 &         {\foiii} &           1F & 119.93 &	 1 &  85.47 &	 1 & 184.58 &	 1  \\
 5006.84 &         {\foiii} &           1F & 361.10 &	 1 & 259.49 &	 1 & 567.04 &	 1  \\
 5411.52 &          {\heii} &          4.7 &	\nd &  \nd &	\nd &  \nd &   5.15 &	 7  \\
 5517.71 &        {\fcliii} &           1F &	\nd &  \nd &	\nd &  \nd &   0.70 &	37  \\
 5666.64 &           {\nii} &            3 &	\nd &  \nd &	\nd &  \nd &   0.54 &	 :  \\
 5679.56 &           {\nii} &            3 &   0.66 &	21 &   0.51 &	 : &   0.85 &	29  \\
 5754.64 &          {\fnii} &           3F &   0.40 &	33 &   0.47 &	 : &   0.85 &	29  \\
 5875.64 &           {\hei} &           11 &  22.06 &	 4 &  22.47 &	 8 &  20.12 &	 6  \\
 6312.10 &         {\fsiii} &           3F &	\nd &  \nd &	\nd &  \nd &   1.00 &	21  \\
 6402.25 &           {\nei} &            1 &	\nd &  \nd &	\nd &  \nd &   0.44 &	 :  \\
 6406.30 &          {\heii} &         5.15 &	\nd &  \nd &	\nd &  \nd &	  * &	 *  \\
 6461.95 &           {\cii} &        17.04 &   0.51 &	25 &	\nd &  \nd &   0.54 &	34  \\
 6527.11 &          {\heii} &         5.14 &	\nd &  \nd &	\nd &  \nd &   0.45 &	39  \\
 6548.03 &          {\fnii} &           1F &   1.40 &	12 &   2.26 &	15 &   5.44 &	 9  \\
 6562.82 &            {\hi} &           H3 & 318.63 &	 7 & 347.18 &	11 & 323.61 &	 8  \\
 6578.05 &           {\cii} &            2 &   0.63 &	21 &   0.40 &	 : &   0.64 &	29  \\
 6583.41 &          {\fnii} &           1F &   4.95 &	 7 &   5.26 &	12 &  15.05 &	 8  \\
 6678.15 &           {\hei} &           46 &   6.27 &	 7 &   6.12 &	13 &   5.72 &	 9  \\
 6683.20 &          {\heii} &         5.13 &	\nd &  \nd &	\nd &  \nd &   0.35 &	 :  \\
 6716.47 &          {\fsii} &           2F &   1.42 &	12 &   0.60 &	32 &   4.32 &	10  \\
 6730.85 &          {\fsii} &           2F &   1.28 &	13 &   0.65 &	30 &   3.45 &	10  \\
 6890.88 &          {\heii} &         5.12 &	\nd &  \nd &	\nd &  \nd &   0.35 &	 :  \\
 7005.67 &          {\farv} &        3P-1D &	\nd &  \nd &	\nd &  \nd &   0.24 &	 :  \\
 7065.28 &           {\hei} &           10 &   2.06 &	10 &   2.30 &	16 &   1.79 &	14  \\
 7135.78 &        {\fariii} &           1F &   5.58 &	 9 &   8.33 &	14 &  13.12 &	10  \\
 7177.50 &          {\heii} &         5.11 &   0.24 &	 : &	\nd &  \nd &   0.71 &	23  \\
 7231.34 &           {\cii} &            3 &	\nd &  \nd &	\nd &  \nd &   0.35 &	 :  \\
 7236.42 &           {\cii} &            3 &	\nd &  \nd &	\nd &  \nd &   0.74 &	22  \\
 7237.17 &           {\cii} &            3 &	\nd &  \nd &	\nd &  \nd &	  * &	 *  \\
 7281.35 &           {\hei} &           45 &	\nd &  \nd &	\nd &  \nd &   0.28 &	 :  \\
 7318.92 &          {\foii} &           2F &	\nd &  \nd &   1.58 &	18 &   3.26 &	12  \\
 7319.99 &          {\foii} &           2F &	\nd &  \nd &	  * &	 * &	  * &	 *  \\
 7329.66 &          {\foii} &           2F &   2.74 &	10 &   1.24 &	20 &   2.73 &	12  \\
 7330.73 &          {\foii} &           2F &	  * &	 * &	  * &	 * &	  * &	 *  \\
 7530.54 &         {\fcliv} &           1F &	\nd &  \nd &	\nd &  \nd &   0.32 &	 :  \\
 7530.57 &           {\cii} &        16.08 &	\nd &  \nd &	\nd &  \nd &	  * &	 *  \\
 7592.74 &          {\heii} &         5.10 &	\nd &  \nd &	\nd &  \nd &   1.04 &	17  \\
 7751.10 &        {\fariii} &           2F &	\nd &  \nd &	\nd &  \nd &   2.76 &	13  \\
 7771.93 &            {\oi} &            1 &	\nd &  \nd &	\nd &  \nd &   1.00 &	18  \\
 7774.17 &            {\oi} &            1 &	\nd &  \nd &	\nd &  \nd &	  * &	 *  \\  
 7775.39 &            {\oi} &            1 &	\nd &  \nd &	\nd &  \nd &	  * &	 *  \\  
 8045.63 &         {\fcliv} &           1F &	\nd &  \nd &	\nd &  \nd &   0.74 &	21  \\
 8196.48 &          {\ciii} &           43 &	\nd &  \nd &	\nd &  \nd &   0.37 &	33  \\
 8236.77 &          {\heii} &          5.9 &   0.72 &	19 &	\nd &  \nd &   1.81 &	15  \\
 8467.25 &            {\hi} &          P17 &	\nd &  \nd &	\nd &  \nd &   0.54 &	24  \\
 8502.48 &            {\hi} &          P16 &	\nd &  \nd &	\nd &  \nd &   0.39 &	30  \\
 8545.38 &            {\hi} &          P15 &	\nd &  \nd &	\nd &  \nd &   0.48 &	26  \\
 8598.39 &            {\hi} &          P14 &	\nd &  \nd &   0.77 &	26 &   0.65 &	22  \\
 8665.02 &            {\hi} &          P13 &	\nd &  \nd &	\nd &  \nd &   1.04 &	18  \\
 8703.25 &        {\nitroi} &            1 &	\nd &  \nd &   0.42 &	37 &	\nd &  \nd  \\ 
 8750.47 &            {\hi} &          P12 &   1.02 &	16 &   1.04 &	24 &   1.13 &	17  \\
 8862.79 &            {\hi} &          P11 &	\nd &  \nd &	\nd &  \nd &   1.44 &	16  \\
 9014.91 &            {\hi} &          P10 &   1.78 &	13 &   1.69 &	22 &   1.75 &	16  \\
 9068.60 &         {\fsiii} &           1F &   8.46 &	12 &  14.67 &	20 &  20.74 &	15  \\
 9229.01 &            {\hi} &           P9 &   1.97 &	13 &   1.97 &	13 &   3.06 &	15  \\
\enddata
\tablecomments{``$\ast$'' indicates that this emission line is blended with nearby ones. 
The total flux of a blend is indicated in the first listed line.
``:'' indicates uncertainties larger than 40\%.}
\end{deluxetable}

\begin{deluxetable}{lccc}
\setlength{\tabcolsep}{0.07in}
\tablecolumns{4}
\tablewidth{0in}
\tablecaption{Extinction, temperatures and densities\label{T-phys1}}
\tablehead{
\colhead{} & \colhead{Abell 46} & \colhead{Abell 63} & \colhead{Ou5} 
}
\startdata
\cb\tablenotemark{a}           & 0.22$\pm$0.08          & 0.55$\pm$0.14        & 0.94$\pm$0.10 \\
\cb\tablenotemark{b}           & 0.04$^{+0.08}_{-0.04}$    & 0.38$\pm$0.11        & 0.77$\pm$0.14 \\
\te\,(\foiii)                  & 12750$\pm$200          &  8750$\pm$950        &   11900$\pm$500       \\
\te\,(\foiii)\tablenotemark{c} & 12050$\pm$200          &  7400$\pm$550        &   10150$\pm$300      \\
\te\,(\fnii)                   & $>$37000               &  $>$24000            &   20000$\pm$8500      \\
\te\,(\fnii)\tablenotemark{c}  & $>$30000               &  $>$22000            &   18800$\pm$5000      \\
\te\,({\hei} 5876/4471)          & 1950$^{+1050}_{-650}$     & 1850$^{+1300}_{-1000}$  &   2800$^{+4500}_{-1100}$\\
\te\,({\hei} 6678/4471)          & 2150$^{+1850}_{-850}$     & 2250$^{+5000}_{-1250}$  &   3200$^{+5000}_{-1500}$\\
\te\,(Balmer jump)             & 1150$\pm$550           & \nd                  &   \nd                \\
\te\,({\oii} 4089/4649)          & 800$^{+2800}_{-800}$      & \nd                  &   $<$1000:           \\
\te\,({\oii} V1/F1)              & 4300$\pm$100           & 4525$\pm$125         &   5020$\pm$360       \\
\elecd\,(\fsii)                & 340$^{+570}_{-220}$       & 600$^{+2100}_{-450}$   &   150$^{+300}_{-100}$   \\
\elecd\,(\foii)                & 2750$^{+880}_{-660}$      & 2600$^{+2500}_{-1300}$  &  1200$^{+550}_{-350}$   \\
\elecd\,(\fariv)               & $<$220:                & \nd                  &   300$^{+800}_{-200}$  \\ 
\elecd\,(CELs, adopted)  & 1590$\pm$600     & 1560$^{+2000}_{-1100}$       &   560$^{+550}_{-300}$  \\ 
\elecd\,(\oii)                 & 2960                 & 1940                  &   3900                 \\
\enddata
\tablecomments{Temperatures are in units of Kelvin, and densities in cm$^{-3}$. ``:'' indicates uncertain values.}
\tablenotetext{a}{Assuming \te=12500~K and \elecd=1000~cm$^{-3}$.}
\tablenotetext{b}{Assuming \te=1000~K and \elecd=1000~cm$^{-3}$.}
\tablenotetext{c}{After correction for the estimated contribution of 
recombination to the auroral lines (see text and Table~\ref{T-recombcontrib}).}
\end{deluxetable}

\clearpage
\begin{deluxetable}{lll}
\setlength{\tabcolsep}{0.07in}
\tablecolumns{3}
\tablewidth{0in}
\tablecaption{Atomic dataset used for collisionally excited lines\label{atomic_cels}}
\tablehead{
\colhead{Ion} & \colhead{Transition probabilities} & \colhead{Collisional strengths}  
}
\startdata
N$^+$    & \citet{froesefischertachiev04}\tnm{a} & \citet{tayal11} \\
O$^+$    & \citet{froesefischertachiev04}        & \citet{kisieliusetal09} \\
O$^{2+}$  & \citet{froesefischertachiev04}\tnm{a} &  \citet{storeyetal14} \\
Ne$^{2+}$ & \citet{galavisetal97}                 & \citet{mclaughlinbell00} \\
S$^+$    & \citet{podobedovaetal09}              & \citet{tayalzatsarinny10} \\
S$^{2+}$  &  \citet{podobedovaetal09}             & \citet{tayalgupta99} \\
Cl$^{2+}$ & \citet{mendoza83}                     & \citet{butlerzeippen89} \\
Cl$^{3+}$ & \citet{kaufmansugar86}                & \citet{galavisetal95} \\
         &  \citet{mendozazeippen82b}            & \\
         &  \citet{ellismartinson84}             & \\
Ar$^{2+}$ & \citet{mendoza83}                     & \citet{galavisetal95} \\
         &  \citet{kaufmansugar86}               & \\
Ar$^{3+}$ & \citet{mendozazeippen82a}             & \citet{ramsbottombell97} \\
Ar$^{4+}$ & \citet{mendozazeippen82b}             & \citet{galavisetal95} \\
          &  \citet{kaufmansugar86}              & \\
          &  \citet{lajohnluke93} & \\
\enddata
\tablenotetext{a}{Adopting the A-values of \citet{2000MNRAS.312..813S} 
results in negligible changes in our \te\ and abundances calculations.}

\end{deluxetable}

\begin{deluxetable}{ll}
\setlength{\tabcolsep}{0.07in}
\tablecolumns{2}
\tablewidth{0in}
\tablecaption{Atomic dataset used for recombination lines\label{atomic_rls}}
\tablehead{
\colhead{Ion} & \colhead{Recombination coefficients}   
}
\startdata
H$^{+}$ &  \citet{sh95}  \\
He$^{+}$ &  \citet{porteretal12, porteretal13}  \\
He$^{2+}$ &  \citet{sh95}  \\
C$^{2+}$ & \citet{daveyetal00}  \\
O$^{2+}$ &  \citet{storey94}  \\
N$^{2+}$ & \citet{fangetal11, fangetal13}  \\
\enddata
\end{deluxetable}

\begin{deluxetable}{lrrr}
\setlength{\tabcolsep}{0.07in}
\tablecolumns{4}
\tablewidth{0in}
\tablecaption{Percentual contribution of recombination to selected CELs\label{T-recombcontrib}}
\tablehead{
\colhead{Line} & 
\colhead{Abell 46} & 
\colhead{Abell 63} & 
\colhead{Ou5} 
}
\startdata
\fnii\,5755       &  13\ph{pp}   &  9\ph{pp}  &   8\ph{p}  \\
\foii\,7320,7330  & 100\ph{pp}   & 56\ph{pp}  & 100\ph{p}  \\
\foiii\,4363      &  14\ph{pp}   & 43\ph{pp}  &  34\ph{p}  \\
\enddata
\end{deluxetable}

\begin{deluxetable}{lrrrrrr}
\setlength{\tabcolsep}{0.07in}
\tablecolumns{7}
\tablewidth{0in}
\tablecaption{Ionic abundances and \adfs\label{T-ionic}}
\tablehead{
\colhead{} & 
\multicolumn{2}{c}{Abell 46} & 
\multicolumn{2}{c}{Abell 63} & 
\multicolumn{2}{c}{Ou5} \\
\cline{2-3} \cline{4-5} \cline{6-7}  \\
\colhead{Ion} & 
\colhead{} & 
\colhead{corrected\tablenotemark{a}} & 
\colhead{} & 
\colhead{corrected\tablenotemark{a}} & 
\colhead{} & 
\colhead{corrected\tablenotemark{a}} 
}
\startdata
He$^+$          	& 11.15\,$\pm$0.02 & 11.15\,$\pm$0.02 & 11.16\,$\pm$0.02      & 11.15\,$\pm$0.03      & 11.14\,$\pm$0.03     & 11.14\,$\pm$0.02	  \\	       
He$^{2+}$        	& 10.35\,$\pm$0.01 & 10.35\,$\pm$0.01 & 10.76\,$\pm$0.01      & 10.75\,$\pm$0.03      & 10.78\,$\pm$0.01     & 10.77\,$\pm$0.01	  \\         
C$^{2+}$ \,(ORLs)&  9.81\,$\pm$0.02 &  9.81\,$\pm$0.02 &  9.57\,$\pm$0.06      &  9.57\,$\pm$0.06      &  9.70\,$\pm$0.05     &  9.70\,$\pm$0.05     \\
N$^+$           	&  5.72\,$\pm$0.04 &  5.78\,$\pm$0.04 &  6.26 $^{+0.15}_{-0.13}$ &  6.48 $^{+0.14}_{-0.09}$ &  6.32\,$\pm$0.05      &  6.49\,$\pm$0.05	  \\          
O$^+$ \,\,(CELs)&  6.83\,$\pm$0.05 &  6.92\,$\pm$0.05 &  7.18 $^{+0.29}_{-0.22}$ &  7.55 $^{+0.22}_{-0.18}$ &  6.97\,$\pm$0.09      &  7.24\,$\pm$0.07	  \\          
O$^+$ \,\,(ORLs)&  \nd             		&  \nd         		    &  \nd      	            &  \nd           		&   8.68\,$\pm$0.07              &  8.66\,$\pm$0.07            \\          
\adf\,(O$^{+}$) 	&  \nd            		 &  \nd            		 &  \nd                  & \nd             			&  {\bf 51}          &  {\bf 26}      \\          
O$^{2+}$ (CELs) &  7.77\,$\pm$0.02 &  7.85\,$\pm$0.02 &  8.17 $^{+0.18}_{-0.14}$ &  8.46 $^{+0.16}_{-0.11}$ &  8.05\,$\pm$0.06      &  8.26\,$\pm$0.05    \\
O$^{2+}$ (ORLs) &  9.93\,$\pm$0.04 &  9.93\,$\pm$0.04 &  9.36 $^{+0.18}_{-0.20}$ &  9.36 $^{+0.18}_{-0.20}$ & 10.01\,$\pm$0.08      & 10.01\,$\pm$0.08     \\
\adf\,(O$^{2+}$)&  {\bf 145}       			&  {\bf 120}       		&  {\bf 15}             & {\bf 8}               &  {\bf 91}             &  {\bf 56}           \\          
Ne$^{2+}$       	&  7.25\,$\pm$0.03 &  7.34\,$\pm$0.03 &  7.70 $^{+0.22}_{-0.18}$ &  8.05 $^{+0.20}_{-0.15}$ &  7.61\,$\pm$0.08      &  7.87\,$\pm$0.05	    \\          
S$^+$          	&  4.81\,$\pm$0.09 &  4.86\,$\pm$0.08 &  4.80\,$\pm$0.24      &  5.01 $^{+0.18}_{-0.16}$ &  5.17\,$\pm$0.07      &  5.32\,$\pm$0.06	    \\          
S$^{2+}$        	&  5.90\,$\pm$0.05 &  5.95\,$\pm$0.05 &  6.47\,$\pm$0.13      &  6.63\,$\pm$0.13      &  6.35\,$\pm$0.08      &  6.48\,$\pm$0.07     \\          
Ar$^{2+}$       	&  5.48\,$\pm$0.04 &  5.53\,$\pm$0.04 &  6.03 $^{+0.15}_{-0.12}$ &  6.23 $^{+0.14}_{-0.11}$ &  5.88\,$\pm$0.05      &  6.03\,$\pm$0.05	     \\          
Ar$^{3+}$       	&  5.16\,$\pm$0.05 &  5.24\,$\pm$0.06 & \nd                   & \nd                   &  5.72\,$\pm$0.08      &  5.94\,$\pm$0.06      \\           
Ar$^{4+}$       	& \nd              & \nd              & \nd                   & \nd                   &  4.48 $^{+0.27}_{-0.38}$ &  4.66 $^{+0.20}_{-0.30}$ \\
Cl$^{2+}$       	& \nd              & \nd              & \nd                   & \nd                   &  4.68 $^{+0.15}_{-0.20}$ &  4.87 $^{+0.17}_{-0.19}$ \\
Cl$^{3+}$      	 & \nd              & \nd              & \nd                   & \nd                   &  4.57\,$\pm$0.10      &  4.72\,$\pm$0.10       \\ 
\enddata
\tablecomments{Abundances are indicated in the usual notation as log(X$^{i+}$/H$^{+}$)+12.}
\tablenotetext{a}{After removal of the estimated contribution of 
recombination to CELs (see text).}
\end{deluxetable}

\begin{deluxetable}{lrrrrrr}
\setlength{\tabcolsep}{0.07in}
\tablecolumns{7}
\tablewidth{0in}
\tablecaption{Total abundances\label{T-totab}}
\tablehead{
\colhead{} & 
\multicolumn{2}{c}{Abell 46} & 
\multicolumn{2}{c}{Abell 63} & 
\multicolumn{2}{c}{Ou5} \\
\cline{2-3} \cline{4-5} \cline{6-7}  \\
\colhead{Element} & 
\colhead{} & 
\colhead{corrected\tablenotemark{a}} & 
\colhead{} & 
\colhead{corrected\tablenotemark{a}} & 
\colhead{} & 
\colhead{corrected\tablenotemark{a}} 
}
\startdata
He & 11.21 $\pm$ 0.01 & 11.22 $\pm$ 0.01 & 11.31 $\pm$ 0.02              & 11.30 $\pm$ 0.02              & 11.30 $\pm$ 0.05              & 11.29 $\pm$ 0.05   \\
N  &  6.72 $\pm$ 0.05 &  6.77 $\pm$ 0.05 &  7.31 $^{+0.15}_{-0.11}$\ph{p\,} &  7.46 $\pm$ 0.11              &  7.46 $\pm$ 0.05              &  7.58 $\pm$ 0.05   \\
O  &  7.85 $\pm$ 0.02 &  7.93 $\pm$ 0.02 &  8.30 $^{+0.20}_{-0.15}$\ph{p\,} &  8.59 $^{+0.16}_{-0.12}$\ph{p\,} &  8.18 $\pm$ 0.06              &  8.40 $\pm$ 0.05   \\
Ne &  7.29 $\pm$ 0.03 &  7.38 $\pm$ 0.03 &  7.78 $^{+0.22}_{-0.18}$\ph{p\,} &  8.14 $^{+0.19}_{-0.15}$\ph{p\,} &  7.70$^{+0.08}_{-0.06}$\ph{pp}   &  7.96 $\pm$ 0.05   \\
S  &  6.15 $\pm$ 0.05 &  6.19 $\pm$ 0.05 &  6.75 $\pm$ 0.13              &  6.89 $\pm$ 0.12              &  6.69 $\pm$ 0.08              &  6.81 $\pm$ 0.07    \\
Ar &  5.67 $\pm$ 0.04 &  5.72 $\pm$ 0.04 &  6.28 $^{+0.14}_{-0.12}$\ph{p\,} &  6.46 $\pm$ 0.12              &  6.15 $\pm$ 0.05              &  6.29 $\pm$ 0.05    \\
Cl &  \nd             & \nd              &  \nd                          &    \nd                        &  5.02 $^{+0.15}_{-0.20}$\ph{p\,} &  5.20 $^{+0.17}_{-0.20}$\ph{p\,}\\
\enddata
\tablecomments{All abundances are from CELs except for He. They are indicated as log(X/H)+12.}
\tablenotetext{a}{After removal of the estimated contribution of recombination to CELs (see text and Table~\ref{T-recombcontrib}).}
\end{deluxetable}


\begin{thebibliography}{}
\bibitem[Afsar \& Ibanoglu(2008)]{a08}
Afsar, M., \& Ibanoglu, C. 2008, \mnras, 391, 802
\bibitem[Balick \& Frank(2002)]{bf02}
Balick, B.,\& Frank, A.  2002, \araa, 40, 439
\bibitem[Barentsen et al.(2014)]{b14}
Barentsen, G., Farnhill, H. J., Drew, J., et al. 2014, MNRAS, 444, 3230
\bibitem[Bear \& Soker(2012)]{bs12}
Bear, E., \& Soker, N. 2012, \apjl, 749, L14
\bibitem[Bear \& Soker(2014)]{bs14}
Bear, E., \& Soker, N. 2014, \mnras, 444, 1698
\bibitem[{Bell, Pollacco \& Hilditch(1994)}]{b94}
Bell, S. A., Pollacco, D. L., \& Hilditch R. W. 1994, \mnras, 270, 449
\bibitem[Bl\"ocker(2001)]{b01}
Bl\"ocker, T. 2001, \apss, 275, 1
\bibitem[Boffin(2014)]{bof14}
Boffin, H. M. J. 2014,  arXiv:1410.3132 
\bibitem[Bond(1980)]{bond80}
Bond, H. E. 1980, IAU Circ. 3480, 0
\bibitem[Bond, Liller \& Mannery(1978)]{bond78}
Bond, H. E., Liller, W., \& Mannery, E. J. 1978, \apj, 223, 252
\bibitem[Butler \& Zeippen(1989)]{butlerzeippen89}
Butler, K., \& {Zeippen}, C.~J.  1989, \aap, 208, 337
\bibitem[Corradi et al.(2011)]{c11}
Corradi, R.L.M., Sabin, L. Miszalski, B., et al. 2011, \mnras, 410, 1349
\bibitem[Corradi et al.(2014)]{c14}
Corradi, R. L. M., Rodr{\'{\i}}guez--Gil, P., Jones, D., et al. 2014, \mnras, 441, 2799
\bibitem[\protect\citeauthoryear{{Davey}, {Storey} \& {Kisielius}}{{Davey}
  et~al.}{2000}]{daveyetal00}
{Davey}, A.~R.,  {Storey}, P.~J.,  \&  {Kisielius}, R.  2000, \aaps, 142, 85
\bibitem[Delgado-Inglada, Morisset \& Stasi\'nska(2014)]{di14}
Delgado-Inglada, G., Morisset, B., \& Stasi\'nska, G. 2014, \mnras,  440, 536
\bibitem[De Marco \& Soker(2002)]{dms02}
De Marco, O., \& Soker, N., 2002, \pasp, 114, 796
\bibitem[De Marco et al.(2011)]{dm11}
De Marco, O., Passy, J.--C., Maxwell, M., et al. 2011, \mnras, 411, 2277
\bibitem[De Marco et al.(2013)]{dm13}
De Marco, O., Passy, J.--C., Frew, D. J., Moe, M., \& Jacoby, G. H. 
2013, \mnras, 428, 2118
\bibitem[Douchin et al.(2014)]{dou14}
Douchin, D., De Marco, O., Frew, D.J., et al. 2014, 
in Asymmetrical Planetary Nebulae VI conference, 
Morisset C. Delgado-Inglada G. \& Torres-Peimbert S. eds., id. 18
(online at \anchor{www.astroscu.unam.mx/apn6/PROCEEDINGS}{www.astroscu.unam.mx/apn6/PROCEEDINGS})
\bibitem[Drew et~al.(2005)]{d05}
Drew, J., Greimel, R., Irwin, M. J., et al. 2005, MNRAS, 362, 753
\bibitem[Ellis \& Martinson(1984)]{ellismartinson84}
Ellis, D.~G., \& Martinson, I.  1984, Phys. Scr., 30, 255
\bibitem[Ercolano et al.(2004)]{e04}
Ercolano, B., Wesson, R., Zhang, Y., et al. 2004, \mnras, 354, 558
\bibitem[{{Esteban} {et~al.}(1998){Esteban}, {Peimbert}, {Torres-Peimbert}, \&
  {Escalante}}]{estebanetal98}
{Esteban}, C., {Peimbert}, M., {Torres-Peimbert}, S., \& {Escalante}, V. 1998,
  \mnras, 295, 401
\bibitem[Esteban et al.(2014)]{e14}
Esteban, C., Garc\'\i a--Rojas, J., Carigi, L., et al. 2014, \mnras, 443, 624
\bibitem[\protect\citeauthoryear{{Fang}, {Storey} \& {Liu}}{{Fang}
  et~al.}{2011}]{fangetal11}
{Fang}, X.,  {Storey}, P.~J., \& {Liu}, X.-W.  2011, \aap, 530, A18
\bibitem[\protect\citeauthoryear{{Fang}, {Storey} \& {Liu}}{{Fang}
  et~al.}{2013}]{fangetal13}
{Fang}, X.,  {Storey}, P.~J., \&  {Liu}, X.-W.  2013, \aap, 550, C2
\bibitem[Farihi et al.(2014)]{f14}
Farihi, J., Wyatt, M.C., Greaves, J.S., et al. 2014, \mnras, 444, 1821
\bibitem[Fitzpatrick(2004)]{f04} 
Fitzpatrick, E. L. 2004, in ``Astrophysics of Dust'', 
Witt A.N., Clayton G.C. \& Draine B.T. eds., ASP Conf. Ser., 
Vol. 309, p. 33
\bibitem[Frankowski \& Soker(2009)]{fs09}
Frankowski, A., \& Soker, N. 2009, New Astronomy, 14, 654
\bibitem[Froese Fischer \& Tachiev(2004)]{froesefischertachiev04}
{Froese Fischer}, C.,  \& {Tachiev}, G.  2004, Atomic Data and Nuclear Data Tables, 87, 1
\bibitem[\protect\citeauthoryear{{Galav\'{\i}s}, {Mendoza} \&
  {Zeippen}}{{Galav\'{\i}s} et~al.}{1995}]{galavisetal95}
{Galav\'{\i}s}, M.~E.,  {Mendoza}, C., \&  {Zeippen}, C.~J.  1995, \aaps, 111, 347
\bibitem[\protect\citeauthoryear{{Galavis}, {Mendoza} \& {Zeippen}}{{Galavis}
  et~al.}{1997}]{galavisetal97}
{Galavis} M.~E.,  {Mendoza} C.,  \& {Zeippen} C.~J.  1997, \aaps, 123, 159
\bibitem[Garc\'\i a--Rojas \& Esteban(2007)]{g07}
Garc\'\i a--Rojas, J., \& Esteban, C., 2007, ApJ, 670, 457
 \bibitem[\protect\citeauthoryear{{Garc{\'{\i}}a-Rojas}, {Pe{\~n}a}, {Morisset},
  {Mesa-Delgado} \& {Ruiz}}{{Garc{\'{\i}}a-Rojas} et~al.}{2012}]{garciarojasetal12}
{Garc{\'{\i}}a-Rojas}, J.,  {Pe{\~n}a}, M.,  {Morisset}, C., 
{Mesa-Delgado}, A.,  \&  {Ruiz}, M.~T.  2012, \aap, 538, A54
\bibitem[\protect\citeauthoryear{{Garc{\'{\i}}a-Rojas}, {Pe{\~n}a}, {Morisset},
  {Delgado-Inglada}, {Mesa-Delgado} \& {Ruiz}}{{Garc{\'{\i}}a-Rojas}
  et~al.}{2013}]{garciarojasetal13}
{Garc{\'{\i}}a-Rojas}, J.,  {Pe{\~n}a}, M.,  {Morisset}, C.,  {Delgado-Inglada},
  G.,  {Mesa-Delgado}, A.,  \&  {Ruiz}, M.~T.  2013, \aap, 558, A122
\bibitem[Gehrz et al.(1998)]{g98}
Gehrz, R.D., Truran, J.W., Williams, R.E., Starrfield, S. 1998, \pasp, 110, 3
\bibitem[Gon\c calves, Corradi, \& Mampaso(2001)]{gcm01}
Gon\c calves, D. R., Corradi, R. L. M., \& Mampaso, A. 2001, \apj, 547, 302  
\bibitem[Hajduk et al.(2005)]{h05}
Hajduk, M., Zijlstra, A.A., Herwig, F. 2005, Science 308, 231
\bibitem[Hamuy et al.(1994)]{h94}
Hamuy, M., Suntzeff, N. B., Heathcote, S. R., et al. 1994, \pasp, 106, 566
\bibitem[Harman \& Seaton(1966)]{hs66}
Harman, R. J., \& Seaton, M. J. 1966, \mnras, 132, 15
\bibitem[Henney \& Stasi\'nska(2010)]{hs10}
Henney, W.J., \& Stasi\'nska, G. 2010, \apj, 711, 881
\bibitem[Kaler \& Jacoby(1989)]{kj89}
Kaler, J. B., \& Jacoby, G. H. 1989, \apj, 345, 871
\bibitem[Kashi \& Soker(2011)]{ks11}
Kashi, A., \& Soker, N. 2011, \mnras, 417, 1466
\bibitem[\protect\citeauthoryear{{Kaufman} \& {Sugar}}{{Kaufman} \&
  {Sugar}}{1986}]{kaufmansugar86}
{Kaufman}, V., \& {Sugar}, J.  1986, Journal of Physical and Chemical Reference
  Data, 15, 321
\bibitem[\protect\citeauthoryear{{Kisielius}, {Storey}, {Ferland} \&
  {Keenan}}{{Kisielius} et~al.}{2009}]{kisieliusetal09}
{Kisielius}, R.,  {Storey}, P.~J.,  {Ferland}, G.~J., \&  {Keenan}, F.~P.  2009, \mnras, 397, 903
\bibitem[Kuruwita(2015)]{ku15}
Kuruwita, R. 2005, Master of Research Thesis, Macquarie University
\bibitem[\protect\citeauthoryear{{LaJohn} \& {Luke}}{{LaJohn} \&
  {Luke}}{1993}]{lajohnluke93}
{LaJohn}, L.,  \& {Luke}, T.~M.  1993, Phys. Scr., 47, 542
\bibitem[Lau, De Marco \& Liu(2011)]{ldl11}
Lau, H. H. B., De Marco, O., \& Liu, X.-W. 2011, \mnras, 410, 1870
\bibitem[Liu(2003)]{l03}
Liu, X.-W. 2003,  in Planetary Nebulae: their evolution and role in the 
Universe, IAU Symp. 209, Kwok, Dopita \& Sutherland eds., p. 339
\bibitem[Liu(2012)]{l12} 
Liu, X.-W. 2012, in IAU Symp. 283, Planetary Nebulae: An Eye to the Future, 
Manchado, A. Stanghellini, L., \& Sch\"onberner, D. eds., p. 131
\bibitem[\protect\citeauthoryear{{Liu}, {Storey}, {Barlow}, {Danziger}, {Cohen}
  \& {Bryce}}{{Liu} et~al.}{2000}]{liuetal00}
{Liu}, X.-W.,  {Storey}, P.~J.,  {Barlow}, M.~J.,  {Danziger}, I.~J.,  {Cohen}, M., \& {Bryce}, M.  2000, \mnras, 312, 585
\bibitem[\protect\citeauthoryear{{Liu}, {Barlow}, {Danziger},\&  {Storey}}{{Liu} et~al.}{2001}]{liuetal01}
{Liu}, X.-W., {Barlow}, M.~J.,  {Danziger}, I.~J., \&  {Storey}, P.~J. 2001, \mnras, 327, 141
\bibitem[Liu et al.(2006)]{l06}
Liu, X.-W., Barlow, M.J., Zhang, Y., Bastin, R.J., Storey P.J., 2006 \mnras, 368, 1959
\bibitem[\protect\citeauthoryear{{Luridiana}, {Morisset} \& {Shaw}}{{Luridiana}
  et~al.}{2014}]{luridianaetal14}
{Luridiana}, V.,  {Morisset}, C.,  \&  {Shaw}, R.~A.  2015, \aap, 573, A42
\bibitem[Lutz et al.(1998)]{lutz98} 
Lutz, J., Alves, D., Becker, A., et al.\ 1998, Bulletin of the American Astronomical Society, 30, 894 
\bibitem[Manick, Miszalski \& McBride (2015)]{ma15}
Manick, R., Miszalski, B., Mc Bride, V. 2015, arXiv:1501.03373
\bibitem[\protect\citeauthoryear{{McLaughlin} \& {Bell}}{{McLaughlin} \&
  {Bell}}{2000}]{mclaughlinbell00}
{McLaughlin}, B.~M., \&  {Bell}, K.~L.  2000, Journal of Physics B Atomic Molecular
  Physics, 33, 597
\bibitem[\protect\citeauthoryear{{Mendoza}}{{Mendoza}}{1983}]{mendoza83}
{Mendoza}, C. 1983, in {Flower} D.~R.,  ed., IAU Symp. 103, Planetary Nebulae
  p.~143
\bibitem[\protect\citeauthoryear{{Mendoza} \& {Zeippen}}{{Mendoza} \&
  {Zeippen}}{1982a}]{mendozazeippen82b}
{Mendoza}, C., \& {Zeippen}, C.~J.  1982a, \mnras, 199, 1025
\bibitem[\protect\citeauthoryear{{Mendoza} \& {Zeippen}}{{Mendoza} \&
  {Zeippen}}{1982b}]{mendozazeippen82a}
{Mendoza}, C.,  \& {Zeippen}, C.~J.  1982b, \mnras, 198, 127
\bibitem[{{Mesa-Delgado} {et~al.}(2012){Mesa-Delgado},
  {N{\'u}{\~n}ez-D{\'{\i}}az}, {Esteban}, {Garc{\'{\i}}a-Rojas},
  {Flores-Fajardo}, {L{\'o}pez-Mart{\'{\i}}n}, {Tsamis}, \&
  {Henney}}]{md12}
{Mesa-Delgado}, A., {N{\'u}{\~n}ez-D{\'{\i}}az}, M., {Esteban}, C., {et~al.}
  2012, \mnras, 426, 614
\bibitem[Miszalski et al.(2009)]{m09}
Miszalski, B., Acker, A., Moffat, A.F.J., Parker, Q. A., Udalski, A. 2009, \aap, 496, 813
\bibitem[Miszalski et al.(2011)]{m11}
Miszalski, B., Jones, D., Rodr\'\i guez-Gil, P., et  al. 2011, \aap,  531, A158
\bibitem[Moe \& De Marco (2001)]{mdm06}
Moe, M., \& De Marco, O.  2006, \apj, 650, 916 
\bibitem[Nicholls et al.(2013)]{n13}
Nicholls, C. P., Melis C., Soszynski I., et al. 2013, \mnras, 431, L33.
\bibitem[Oke(1990)]{oke90}
{Oke}, J.~B., 1990, \aj, 99, 1621
\bibitem[Parson et al.(2014)]{par14}
Parson, S. G., Marsch, T. R., Bours, M. C. P., et al. 2014, \mnras, 438, L91
\bibitem[Passy et al.(2012)]{pa12}
Passy, J.--C., De Marco, O., Fryer, C. L.,  et al. 2012, \apj, 744, 52
\bibitem[{{Peimbert} {et~al.}(2005){Peimbert}, {Peimbert}, \& {Ruiz}}]{apeimbertetal05}
{Peimbert}, A., {Peimbert}, M., \& {Ruiz}, M.~T. 2005, ApJ, 634, 1056
\bibitem[\protect\citeauthoryear{{Peimbert}, {Peimbert}, {Delgado-Inglada}, {Garc\'{i}a-Rojas} \& {Pe\~na}}{{Peimbert}
  et~al.}{2014}]{peimbertetal14}
{Peimbert}, A.,  {Peimbert}, M., {Delgado-Inglada}, G., {Garc\'{\i}a-Rojas}, J., \&  {Pe\~na}, M.  2014, Rev. Mex. Astron. Astrofis., 50, 329
\bibitem[\protect\citeauthoryear{{Podobedova}, {Kelleher} \&
  {Wiese}}{{Podobedova} et~al.}{2009}]{podobedovaetal09}
{Podobedova}, L.~I.,  {Kelleher}, D.~E.,  \&  {Wiese}, W.~L.  2009, Journal of
  Physical and Chemical Reference Data, 38, 171
\bibitem[Pollacco \& Bell(1997)]{pb97}
Pollacco, D.L., \& Bell, S.A. 1997, \mnras, 284, 32
\bibitem[\protect\citeauthoryear{{Porter}, {Ferland}, {Storey} \&
  {Detisch}}{{Porter} et~al.}{2012}]{porteretal12}
{Porter}, R.~L.,  {Ferland}, G.~J.,  {Storey}, P.~J., \& {Detisch}, M.~J.  2012, \mnras, 425, L28
\bibitem[\protect\citeauthoryear{{Porter}, {Ferland}, {Storey} \&
  {Detisch}}{{Porter} et~al.}{2013}]{porteretal13}
{Porter}, R.~L.,  {Ferland}, G.~J.,  {Storey}, P.~J.,  \& {Detisch}, M.~J.  2013, \mnras, 433, L89
\bibitem[Pottasch(1984)]{pot84}
Pottasch, S.R. 1984, {\it `Planetary nebulae'}, D.Reidel, Dordrecht
\bibitem[\protect\citeauthoryear{{Ramsbottom} \& {Bell}}{{Ramsbottom} \&
  {Bell}}{1997}]{ramsbottombell97}
{Ramsbottom} C.~A., \& {Bell} K.~L. 1997, Atomic Data and Nuclear Data Tables, 66, 65
\bibitem[Ricker \& Tamm(2012)]{rt12}
Ricker, P. M., \& Tamm, R. E. 2012, \apj, 746, 74
\bibitem[Ruiz et al.(2003)]{ruizetal03}
Ruiz, M. T.,  Peimbert, A., Peimbert, M., \& Esteban, C. 2003, \apj, 595, 247
\bibitem[Sandquist et al.(1998)]{sand98}
Sandquist, E. L., Taam, R. E., Chen, X., Bodenheimer, P., \& 
Burkert, A. 1998, \apj, 500, 909
\bibitem[Schaub, Bodman \& Hillwig(2012)]{s12}
Schaub, S. C., Bodman, E., \& Hillwig, T. C. 2012, JSARA, 5, 2
\bibitem[Schlegel et al.(1998)]{1998ApJ...500..525S} Schlegel, D.~J., 
Finkbeiner, D.~P., \& Davis, M.\ 1998, \apj, 500, 525 
\bibitem[Soker(1997)]{s97}
Soker, N. 1997, \apjs, 112, 487
\bibitem[Soker(2006)]{s06}
Soker, N. 2006, \apj, 645, L57
\bibitem[\protect\citeauthoryear{{Storey}}{{Storey}}{1994}]{storey94}
Storey, P.~J.  1994, \aap, 282, 999
\bibitem[Storey \& Hummer(1995)]{sh95}
Storey, P. J., \& Hummer, D. G. 1995, \mnras, 272, 41
\bibitem[Storey \& Zeippen(2000)]{2000MNRAS.312..813S} 
Storey, P.~J., \& Zeippen, C.~J.\ 2000, \mnras, 312, 813 
\bibitem[\protect\citeauthoryear{{Storey}, {Sochi} \& {Badnell}}{{Storey}
  et~al.}{2014}]{storeyetal14}
{Storey}, P.~J.,  {Sochi}, T., \&  {Badnell}, N.~R.  2014, \mnras, 441, 3028
\bibitem[\protect\citeauthoryear{{Tayal}}{{Tayal}}{2011}]{tayal11}
{Tayal}, S.~S. 2011, \apjs, 195, 12
\bibitem[\protect\citeauthoryear{{Tayal} \& {Gupta}}{{Tayal} \&
  {Gupta}}{1999}]{tayalgupta99}
{Tayal}, S.~S.,  \& {Gupta} G.~P.  1999, \apj, 526, 544
\bibitem[\protect\citeauthoryear{{Tayal} \& {Zatsarinny}}{{Tayal} \&
  {Zatsarinny}}{2010}]{tayalzatsarinny10}
{Tayal}, S.~S., \& {Zatsarinny}, O.  2010, \apjs, 188, 32
\bibitem[{{Tsamis} {et~al.}(2003){Tsamis}, {Barlow}, {Liu}, {Danziger}, \&
  {Storey}}]{tsamisetal03}
{Tsamis}, Y.~G., {Barlow}, M.~J., {Liu}, X.-W., {Danziger}, I.~J., \& {Storey},
  P.~J. 2003, \mnras, 338, 687
\bibitem[Tsamis et al.(2004)]{t04}
Tsamis, Y.~G., Barlow, M.~J., Liu, X.-W., Storey, P.~J., \& Danziger, 
I.~J. 2004, \mnras, 353, 953
\bibitem[Tsamis et al.(2008)]{t08}
Tsamis, Y.~G., Walsh, J.~R., P\'equignot, D.,  et al. 2008, \mnras, 386, 22
\bibitem[Tsamis et al.(2011)]{t11} 
Tsamis, Y.~G., Walsh, J.~R., V{\'{\i}}lchez, J.~M., \& P{\'e}quignot, D.\ 2011, \mnras, 412, 1367 
\bibitem[Tsamis et al.(2013)]{t13} 
Tsamis, Y.~G., Flores-Fajardo, N., Henney, W.~J., Walsh, J.~R., \& Mesa-Delgado, A.\ 2013, \mnras, 430, 3406 
\bibitem[Yaron et al.(2005)]{y05}
Yaron, O., Prialnik, D., Shara, M. M., \& Kovetz, A. 2005, \apj, 623, 398
\bibitem[Wesson, Liu, \& Barlow(2003)]{w03}
Wesson, R., Liu, X.-W., \& Barlow, M.~J. 2003, \mnras, 340, 253
\bibitem[Wesson et al.(2008)]{w08}
Wesson, R.,  Barlow, M.~J., Liu, X.-W., et al. 2008, \mnras, 383, 1639
\bibitem[Zhang et al.(2005)]{zhangetal05}
Zhang, Y., Liu, X.-W., Liu, Y., \& Rubin, R.~H. 2005, \mnras, 358, 457
\bibitem[Zorotovic \& Schreiber(2013)]{z13}
Zorotovic, M., \& Schreiber, M.~R. 2013, \aap, 549, A95
\end{thebibliography}
\end{document}